\documentclass[iop,apj,twocolumn,twocolappendix,numberedappendix]{emulateapj}
\usepackage{apjfonts}
\usepackage{multirow}
\usepackage{textcomp}
\usepackage{amsmath}
\usepackage{microtype}
\usepackage{subfigure}
\usepackage{xcolor,ulem}
\usepackage[breaklinks=true]{hyperref}

\begin{document}

\title{Parameter estimation on gravitational waves from neutron-star binaries with spinning components}

\slugcomment{Submitted to ApJ}

\author{Ben~Farr\altaffilmark{1}}
\email{farr@uchicago.edu}
\author{Christopher~P.~L.~Berry\altaffilmark{2}}

\author{Will~M.~Farr\altaffilmark{2}}
\author{Carl-Johan~Haster\altaffilmark{2}}
\author{Hannah~Middleton\altaffilmark{2}}

\author{Kipp~Cannon\altaffilmark{3}}
\author{Philip~B.~Graff\altaffilmark{4,5}}
\author{Chad~Hanna\altaffilmark{6}}
\author{Ilya~Mandel\altaffilmark{2}} 
\author{Chris~Pankow\altaffilmark{7}}
\author{Larry~R.~Price\altaffilmark{8}}
\author{Trevor~Sidery\altaffilmark{2}}
\author{Leo~P.~Singer\altaffilmark{9}}
\author{Alex~L.~Urban\altaffilmark{7}} 
\author{Alberto~Vecchio\altaffilmark{2}} 
\author{John~Veitch\altaffilmark{2}} 
\author{Salvatore~Vitale\altaffilmark{10}}

\altaffiltext{1}{Enrico Fermi Institute and Kavli Institute for Cosmological Physics, University of Chicago, Chicago, IL 60637, USA}
\altaffiltext{2}{School of Physics \& Astronomy, University of Birmingham, Birmingham, B15 2TT, UK}
\altaffiltext{3}{Canadian Institute for Theoretical Astrophysics, 60 St.\ George Street, University of Toronto, Toronto, Ontario, M5S 3H8, Canada}
\altaffiltext{4}{Department of Physics, University of Maryland--College Park, College Park, MD 20742, USA}
\altaffiltext{5}{Gravitational Astrophysics Lab, NASA Goddard Space Flight Center, Greenbelt, MD 20771, USA}
\altaffiltext{6}{The Pennsylvania State University, University Park, PA 16802, USA}
\altaffiltext{7}{Leonard E.\ Parker Center for Gravitation, Cosmology, and Astrophysics, University of Wisconsin--Milwaukee, Milwaukee, WI 53201, USA}
\altaffiltext{8}{LIGO Laboratory, California Institute of Technology, Pasadena, CA 91125, USA}
\altaffiltext{9}{Astrophysics Science Division, NASA Goddard Space Flight Center, Code 661, Greenbelt, MD 20771, USA}
\altaffiltext{10}{Massachusetts Institute of Technology, 185 Albany St, Cambridge, MA 02139, USA}

\shorttitle{PE on GWs from BNSs with spinning components}
\shortauthors{Farr et al.}

\begin{abstract}
Inspiraling binary neutron stars are expected to be one of the most significant sources of gravitational-wave signals for the new generation of advanced ground-based detectors. We investigate how well we could hope to measure properties of these binaries using the Advanced LIGO detectors, which began operation in September 2015. We study an astrophysically motivated population of sources (binary components with masses $1.2~\mathrm{M}_\odot$--$1.6~\mathrm{M}_\odot$ and spins of less than $0.05$) using the full LIGO analysis pipeline. While this simulated population covers the observed range of potential binary neutron-star sources, we do not exclude the possibility of sources with parameters outside these ranges; given the existing uncertainty in distributions of mass and spin, it is critical that analyses account for the full range of possible mass and spin configurations. We find that conservative prior assumptions on neutron-star mass and spin lead to average fractional uncertainties in component masses of $\sim 16\%$, with little constraint on spins (the median $90\%$ upper limit on the spin of the more massive component is $\sim 0.7$).  Stronger prior constraints on neutron-star spins can further constrain mass estimates, but only marginally.  However, we find that the sky position and luminosity distance for these sources are not influenced by the inclusion of spin; therefore, if LIGO detects a low-spin population of BNS sources, less computationally expensive results calculated neglecting spin will be sufficient for guiding electromagnetic follow-up.
\end{abstract}


\section{Introduction}

As we enter the advanced-detector era of ground-based gravitational-wave (GW) astronomy, it is critical that we understand the abilities and limitations of the analyses we are prepared to conduct. Of the many predicted sources of GWs, binary neutron star (BNS) coalescences are paramount; their progenitors have been directly observed \cite{Lorimer_2008}, and the advanced detectors will be sensitive to their GW emission up to $\sim 400~\mathrm{Mpc}$ away \citep{2013arXiv1304.0670L}.

When analyzing a GW signal from a circularized compact binary merger, strong degeneracies exist between parameters describing the binary (e.g., distance and inclination). To properly estimate any particular parameter(s) of interest, the marginal distribution is estimated by integrating the joint posterior probability density function (PDF) over all other parameters. In this work we sample the posterior PDF using software implemented in the \textsc{LALInference} library \citep{Veitch_2014}, specifically we use results from \textsc{LALInfernce\_nest} \linebreak \citep{Veitch_2010}, a nest sampling algorithm \citep{Skilling2006}, and \textsc{LALInference\_MCMC} \citep{Christensen_2003,R_ver_2006,van_der_Sluys_2008}, a Markov-chain Monte Carlo algorithm \citep[chapter 12]{Gregory2005}.

Previous studies of BNS signals have largely assessed parameter constraints assuming negligible neutron-star (NS) spin, restricting models to nine parameters. This simplification has largely been due to computational constraints, but the slow spin of NSs in short-period BNS systems observed to date \citep[e.g.,][]{Mandel_2010} has also been used for justification. However, proper characterization of compact binary sources \textit{must} account for the possibility of non-negligible spin, otherwise parameter estimates will be biased \citep{Buonanno_2009,Berry_2014}.  This bias can potentially lead to incorrect conclusions about source properties, and even misidentification of source classes.

Numerous studies have looked at the BNS parameter estimation abilities of ground-based GW detectors such as the Advanced Laser Interferometer Gravitational-Wave Observatory \citep[aLIGO;][]{Aasi_2015} and Advanced Virgo \citep[AdV;][]{Acernese_2014} detectors. \citet{Nissanke_2010,Nissanke_2011} assessed localization abilities on a simulated non-spinning BNS population. \citet{Veitch_2012} looked at several potential advanced-detector networks and quantified the parameter-estimation abilities of each network for a signal from a fiducial BNS with non-spinning NSs. \citet{Aasi_2013} demonstrated the ability to characterize signals from non-spinning BNS sources with waveform models for spinning sources using Bayesian stochastic samplers in the \textsc{LALInference} library \citep{Veitch_2014}.  \citet{Hannam_2013} used approximate methods to quantify the degeneracy between spin and mass estimates, assuming the compact objects' spins are aligned with the orbital angular momentum of the binary \citep[but see][]{Haster_2015}. \citet{Rodriguez_2014} simulated a collection of loud signals from non-spinning BNS sources in several mass bins and quantified parameter estimation capabilities in the advanced-detector era using non-spinning models.  \citet{Chatziioannou_2014} introduced precession from spin--orbit coupling and found that the additional richness encoded in the waveform could reduce the mass--spin degeneracy, helping BNSs to be distinguished from NS--black hole (BH) binaries; \citet{Littenberg:2015tpa} conducted a similar analysis of a large catalog of sources and found that it is difficult to infer the presence of a mass gap between NSs and BHs \citep{Ozel:2010su,Farr:2010tu,Kreidberg:2012ud}, although this may still be possible using a population of a few tens of detections \citep{Mandel:2015spa}.  Finally, \citet{Singer_2014} and the follow-on \citet{Berry_2014} represent an (almost) complete end-to-end simulation of BNS detection and characterization during the first $1$--$2$ years of the advanced-detector era. These studies simulated GWs from an astrophysically motivated BNS population, then detected and characterized sources using the search and follow-up tools that are used for LIGO--Virgo data analysis \citep{WhitePaper2014,TheLIGOScientific:2016wfe}.   The final stage of the analysis missing from these studies is the computationally expensive characterization of sources while accounting for the compact objects' spins and their degeneracies with other parameters.  The present work is the final step of BNS characterization for the \citet{Singer_2014} simulations using waveforms that account for the effects of NS spin.

We begin with a brief introduction to the source catalog used for this study and \citet{Singer_2014} in section \ref{sec:sources}. Then, in section \ref{sec:spin} we describe the results of parameter estimation from a full analysis that includes spin. In section \ref{sec:mass} we look at mass estimates in more detail, and spin-magnitude estimates in section \ref{sec:spin-magnitudes}. In section \ref{sec:extrinsic} we consider the estimation of extrinsic parameters: sky position (section \ref{sec:sky}) and distance (section \ref{sec:distance}), which we do not expect to be significantly affected by the inclusion of spin in the analysis templates. We summarize our findings in section \ref{sec:conclusions}. A comparison of computational costs for spinning and non-spinning parameter estimation is given in appendix \ref{ap:CPU}.

\section{Source Simulation and Selection}\label{sec:sources}

We have restricted our study to the first year of the advanced-detector era, using the same $250$ simulations that \citet{Singer_2014} analysed with non-spinning parameter estimation. For these, Gaussian noise was generated using the `early' 2015 aLIGO noise curve found in \citet{Barsotti:2012}. Approximately $50,000$ BNS sources were simulated, using the SpinTaylorT4 waveform model \citep{Buonanno_2003,Buonanno_2009}, a post-Newtonian inspiral model that includes the effects of precession, to generate the GW signals. Component masses were uniformly distributed between $1.2~\mathrm{M}_\odot$ and $1.6~\mathrm{M}_\odot$, which reflects the range of observed BNS masses \citep{_zel_2012}. Component spins were isotropically oriented, with magnitudes $\chi_{1,\,2} = c |\boldsymbol{S}_{1,\,2}|/G m_{1,\,2}^2$ drawn uniformly between $0$ and $0.05$; here $|\boldsymbol{S}_{1,\,2}|$ are the NSs' spin angular momenta and $m_{1,\,2}$ their mass, the indices $1$ and $2$ correspond to the more and less massive components of the binary, respectively.  The range of simulated spin magnitudes was chosen to be consistent with the observed population of short-period BNS systems, currently bounded by PSR J0737$-$3039A \citep{Burgay_2003,Brown_2012} from above.  Finally, sources were distributed uniformly in volume (i.e. uniform in distance cubed) to a maximum distance at which the loudest signal would produce a network signal-to-noise ratio (S/N) of $\rho_\mathrm{net} = 5$ \citep{Singer_2014}, where $\rho_\mathrm{net} = \sum_i \rho_i^2$ is the individual detector S/Ns $\rho_i$ combined in quadrature.

Of this simulated population, detectable sources were selected using the \textsc{gstlal\_inspiral} matched-filter detection pipeline \citep{Cannon_2012} with a single-detector S/N threshold $\rho>4$ and false alarm rate (FAR) threshold of $\mathrm{FAR}<10^{-2}~\mathrm{yr}^{-1}$.  The FAR for real detector noise is largely governed by non-stationary noise transients in the data that can mimic GWs from compact binary mergers, which \citet{Berry_2014} demonstrate make negligible difference to parameter estimation for the (low-FAR, BNS) signals considered here.  Because our simulated noise is purely stationary and Gaussian with no such artifacts, FAR estimates are overly optimistic. To compensate, an additional threshold on the network S/N of $\rho_\mathrm{net} > 12$ was applied. This S/N threshold is consistent with the above FAR threshold when applied to data similar to previous science runs \citep{2013arXiv1304.0670L,Berry_2014}. A random subsample of $250$ detections were selected for parameter estimation with \textsc{LALInference}.\footnote{The mean (median) $\rho_\mathrm{net}$ of the set of $250$ events is $16.7$ ($14.6$).} This mass and spin distributions of this subset is statistically consistent with those the sources were drawn from \citep{Berry_2014}. See \citet{Singer_2014} for more details regarding the simulated data and \textsc{gstlal\_inspiral} analyses.

\section{Spinning Analysis}
\label{sec:spin}

\citet{Singer_2014} details the detection, low-latency localization, and medium-latency (i.e.\ non-spinning) follow-up of the simulated signals in 2015. In this work we perform the expensive task of full parameter estimation that accounts for non-zero compact-object spin. Whereas \citet{Singer_2014} used the (non-spinning) TaylorF2 waveform model, we make use of the SpinTaylorT4 waveform model \citep{Buonanno_2003,Buonanno_2009}, parameterized by the fifteen parameters that uniquely define a circularized compact binary inspiral.\footnote{The fifteen parameters are two masses (either component masses or the chirp mass and mass ratio); six spin parameters describing the two spins (magnitudes and orientations); two coordinates for sky position; distance; an inclination angle; a polarization angle; a reference time, and the orbital phase at this time \citep[see][for more details]{Veitch_2014}. The masses and spins are intrinsic parameters which control the evolution of the binary, the others are extrinsic parameters which describe its orientation and position.}

We assume the objects to be point masses with no tidal interactions.  The estimation of tidal parameters using post-Newtonian approximations is rife with systematic uncertainties that are comparable in magnitude to statistical uncertainties \citep{Yagi_2014,Wade_2014}. Though marginalizing over uncertainties in tidal parameters can affect estimates of other parameters, the fact that tidal interactions only impact the evolution of the binary at late times (only having a measurable impact at frequencies above $\sim450~\mathrm{Hz}$; \citealt{Hinderer_2010}) limits both their measurability and the resulting biases in other parameter estimates caused by ignoring them \cite{Damour_2012}.

The simulated population of BNS systems contains slowly spinning NSs, with masses between $1.2~\mathrm{M}_\odot$ and $1.6~\mathrm{M}_\odot$ and spin magnitudes $\chi < 0.05$.  This choice was motivated by the characteristics of NSs found thus far in Galactic BNS systems expected to merge within a Hubble time through GW emission. However, NSs \textit{outside} of BNS systems have been observed with spins as high as $\chi = 0.4$ \citep{Hessels_2006,Brown_2012}, and depending on the NS equation of state (EOS) could theoretically have spins as high as $\chi \lesssim 0.7$ \citep{Lo_2011} without breaking up.  For these reasons, the prior assumptions used for Bayesian inference of source parameters are more broad than the spin range of the simulated source population.

To simulate a real analysis scenario where the class of compact binary and the NS EOS are not known, we use uniform priors in component masses between $0.6~\mathrm{M}_\odot$ and $5.0~\mathrm{M}_\odot$ to avoid any prior constraints on mass posteriors, and our standard BH spin prior: uniform in spin magnitudes $\chi_{1,\,2} \sim U(0, 1)$ and isotropic in spin orientation. Prior distributions for the location and orientation of the binary match that of the simulated population, i.e.\ isotropically oriented and uniform in volume (out to a maximum distance of $218.9~\mathrm{Mpc}$, safely outside the detection horizon, which is $\sim137~\mathrm{Mpc}$ for a $1.6~\mathrm{M}_\odot$--$1.6~\mathrm{M}_\odot$ binary).\footnote{The mean (median) true distance for the set of $250$ events is $52.1~\mathrm{Mpc}$ ($47.8~\mathrm{Mpc}$), and the maximum is $124.8~\mathrm{Mpc}$.}  Choosing any particular upper bound for spin magnitude would require either assuming hard constraints on NS spin-up, which are based upon observations with hard-to-quantify selection effects, or making assumptions regarding the unknown EOS of NSs. For these reasons we choose not to rule out compact objects with high spin a priori by using an upper limit of $\chi < 1$, encompassing all allowed NS and BH spins.  In section \ref{subsec:prior_constraints} we look at more constraining spin priors, and particularly how such choices can affect mass estimates.

We describe parameter-estimation accuracy using several different quantities, depending upon the parameter of interest.
\begin{itemize}
\item Simplest is the fractional uncertainty $\sigma_x/\langle x\rangle$, where $\sigma_x$ and $\langle x\rangle$ are the standard deviation and mean of the distributions for parameter $x$ respectively. This is particularly useful for showing how uncertainty scales with S/N: in the limit of high S/N, the standard deviation can be approximated from the (inverse) Fisher matrix and scales inversely with the S/N \citep{Vallisneri_2008}.
\item The credible interval $\mathrm{CI}_p^{x}$ in the range that contains the central $p$ of the integrated posterior, with $(1-p)/2$ falling both above and below the limits \citep{Aasi_2013}. Specifying the credible interval for several values of $p$ gives information about the shape of the posterior.
\item As an alternative to credible intervals, we use credible upper or lower bounds. These are the one-sided equivalents of credible intervals, and are useful for distributions that are peaked towards one end of the parameter range or for parameters we are interested in putting a limit upon (the spin magnitude satisfies both of these criteria).
\item For sky-localization, we use credible regions (the two-dimensional generalization of the credible interval) which are the smallest sky area that encompasses a given total posterior probability. The credible region for a total posterior probability $p$ is defined as
\begin{equation}
\mathrm{CR}_p = \underset{A}{\arg\!\max} \int_A \mathrm{d}\boldsymbol{\Omega} P_{\Omega}(\boldsymbol{\Omega}),
\label{eq:CR}
\end{equation}
where $P_{\Omega}(\boldsymbol{\Omega})$ is the posterior PDF over sky position $\boldsymbol{\Omega}$, and $A$ is the sky area integrated over \citep{Sidery_2014}. We also consider the searched area $A_\ast$, the area of the smallest credible region that includes the true location.
\end{itemize}
  
To check that differences between our spinning and non-spinning analyses were a consequence of the inclusion of spin and not because of a difference between waveform approximants, we also ran SpinTaylorT4 analyses with spins fixed to $\chi_{1,~2}=0$. There were no significant differences in parameter estimation between the non-spinning TaylorF2 and zero-spin SpinTaylorT4 results for any of the quantities we examined.\footnote{Using as an example the chirp mass, the most precisely inferred parameter, we can compare the effects of switch from a non-spinning to a spinning analysis to those from switching waveform approximants by comparing the difference the posterior means $\langle \mathcal{M}_\mathrm{c}\rangle$. The difference between means from the SpinTaylorT4 analyses with and without spin, is an order of magnitude greater than the difference between the zero-spin SpinTaylorT4 and TaylorF2 analyses: defining the log-ratio $\xi = \log_{10}(|\langle \mathcal{M}_\mathrm{c}\rangle^\mathrm{S} - \mathcal{M}_\mathrm{c}\rangle^0|/|\langle \mathcal{M}_\mathrm{c}\rangle^\mathrm{NS} - \mathcal{M}_\mathrm{c}\rangle^0|)$, where the superscripts $\mathrm{S}$, $0$ and $\mathrm{NS}$ indicates results of the fully spinning SpinTaylorT4, the zero-spin SpinTaylorT4 and the non-spinning TaylorF2 analyses respectively, the mean (median) value of $\xi$ is $0.90$ ($1.04$), and $92.4\%$ of events have $\xi > 0$ (indicating that the shift in the mean from introducing spin is larger than the shift from switching approximants).} Therefore, we only use the TaylorF2 results to illustrate the effects of neglecting spin.

\subsection{Mass Estimates}\label{sec:mass}
To maximize sampling efficiency, model parameterizations are chosen to minimize degeneracies between parameters.  To leading order, the post-Newtonian expansion of the waveform's phase evolution depends on the \textit{chirp mass}, $\mathcal{M}_\mathrm{c} = (m_1 m_2)^{3/5} (m_1 + m_2)^{-1/5}$, making it a \textit{very} well constrained parameterization of binary mass.  The second mass parameter used is the mass ratio $q = m_2/m_1$, where $0 < q \leq 1$.  Detectors are much less sensitive to the mass ratio, and strong degeneracies with spin make constraints on $q$ even worse \citep{Cutler_1994}.  It is primarily the uncertainty in $q$ that governs the uncertainty in component masses $m_1$ and $m_2$.

Figure \ref{fig:mass_pdfs} shows the superimposed, one-dimensional marginal posterior PDFs and cumulative density functions (CDFs) for the chirp mass (centered on each mean) and mass ratio for all $250$ events.  As a representation of a typical event's posterior distribution, we show the average PDFs and CDFs, where the average is taken over all $250$ posterior PDFs and CDFs at each point. Chirp-mass distributions are usually well approximated by normal distributions about the mean, while mass ratio estimates have broad support across most of the prior range, whereas the simulated population had a narrower range between $0.75$ and $1$.

To trace individual parameter uncertainties across the population we use the fractional uncertainties in chirp mass $\sigma_{\mathcal{M}_\mathrm{c}}/\langle\mathcal{M}_\mathrm{c}\rangle$ and mass ratio $\sigma_q/\langle q\rangle$. The chirp mass and mass ratio conveniently cover mass space (which is why they are used for sampling), but the total mass $M = m_1 + m_2$ is also of interest for determining the end product of the merger, so we also plot the fractional uncertainty $\sigma_M/\langle M\rangle$. The mean (median) fractional uncertainties in chirp mass, mass ratio and total mass for the simulated population are $0.0676\%$ ($0.0642\%$), $28.7\%$ ($28.4\%$) and $6.15\%$ ($5.81\%$) respectively. For comparison, the mean (median) fractional uncertainties in chirp mass, mass ratio and total mass from the non-spinning analysis are $0.0185\%$ ($0.0165\%$), $8.90\%$ ($8.79\%$) and $0.542\%$ ($0.491\%$) respectively.  We further examine the impact of spin on mass measurements in section \ref{subsec:prior_constraints}.

The fractional uncertainties for the chirp mass, mass ratio and total mass all decrease as S/N increases, as shown in Figure \ref{fig:Mc_q_std_snr}, which also shows results from the non-spinning analysis. As expected from Fisher-matrix studies \citep[e.g.,][]{FinnChernoff}, most appear to be inversely proportional to the S/N: the exception is $\sigma_q/\langle q\rangle$ from the spinning analysis, which better fit as $\propto \rho_\mathrm{net}^{-1/2}$. We do not suspect there is anything fundamental about the $\propto \rho_\mathrm{net}^{-1/2}$, rather it is a useful rule-of-thumb. The behaviour can still be understood from a Fisher-matrix perspective, which predicts a Gaussian probability distribution (with width $\propto \rho_\mathrm{net}^{-1}$). Since the mass ratio is constrained to be $0 \leq q \leq 1$, if the width of a Gaussian is large, it is indistinguishable from a uniform distribution and the standard deviation tends to a constant $1/\sqrt{12} \simeq 0.289$. When the width of the Gaussian is small ($\lesssim 0.1$), the truncation of the distribution is negligible and the standard deviation behaves as expected, as is the case for the non-spinning results. The standard deviations obtained for the spinning runs lie in the intermediate regime, between being independent of S/N and scaling inversely with it \citep[cf.][]{Littenberg:2015tpa}; the mean (median) standard deviation $\sigma_q$ is $0.182$ ($0.183$).\footnote{The uncertainty  for the symmetric mass ratio $\eta = m_1m_2/(m_1 + m_2)^2$, which is constrained to be $0 \leq \eta \leq 1/4$, does scale approximately as $\rho_\mathrm{net}^{-1}$. The mean (median) standard deviation $\sigma_\eta$ for the spinning runs is $2.00\times 10^{-2}$ ($1.95\times 10^{-2}$).} The mass--spin degeneracy broadens the posteriors for both the chirp mass, the mass ratio and the total mass; a consequence of the broadening for the mass ratio is that the uncertainty does not decrease as rapidly with S/N (over the range considered here).
  
Projecting the tightly constrained chirp mass and poorly constrained mass ratio $90\%$ credible region from $\mathcal{M}_\mathrm{c}$--$q$ space into component-mass space makes it obvious how important mass-ratio uncertainties are for extracting astrophysical information.  The credible regions in component-mass space are narrow bananas that lie along lines of constant chirp mass, bounded by the constraints on mass ratio (see Figure \ref{fig:comp_masses} for some examples posteriors).

\begin{figure*}
  \centering
  \includegraphics[width=1.6\columnwidth]{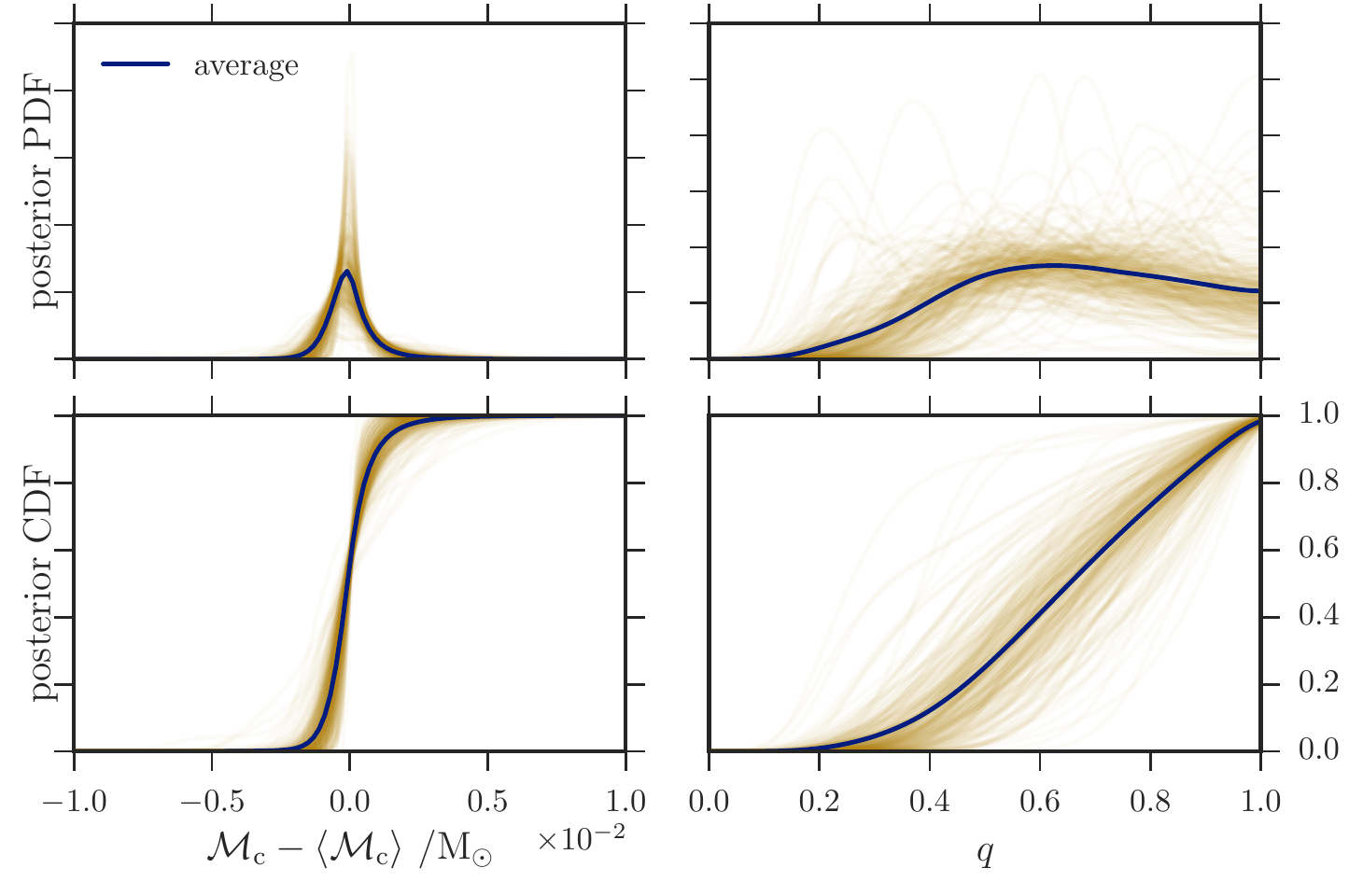}
  \caption{\protect\input{mass_pdf_cdf-caption.tex}}
\end{figure*}

\begin{figure}
  \centering
  \includegraphics[width=0.85\columnwidth]{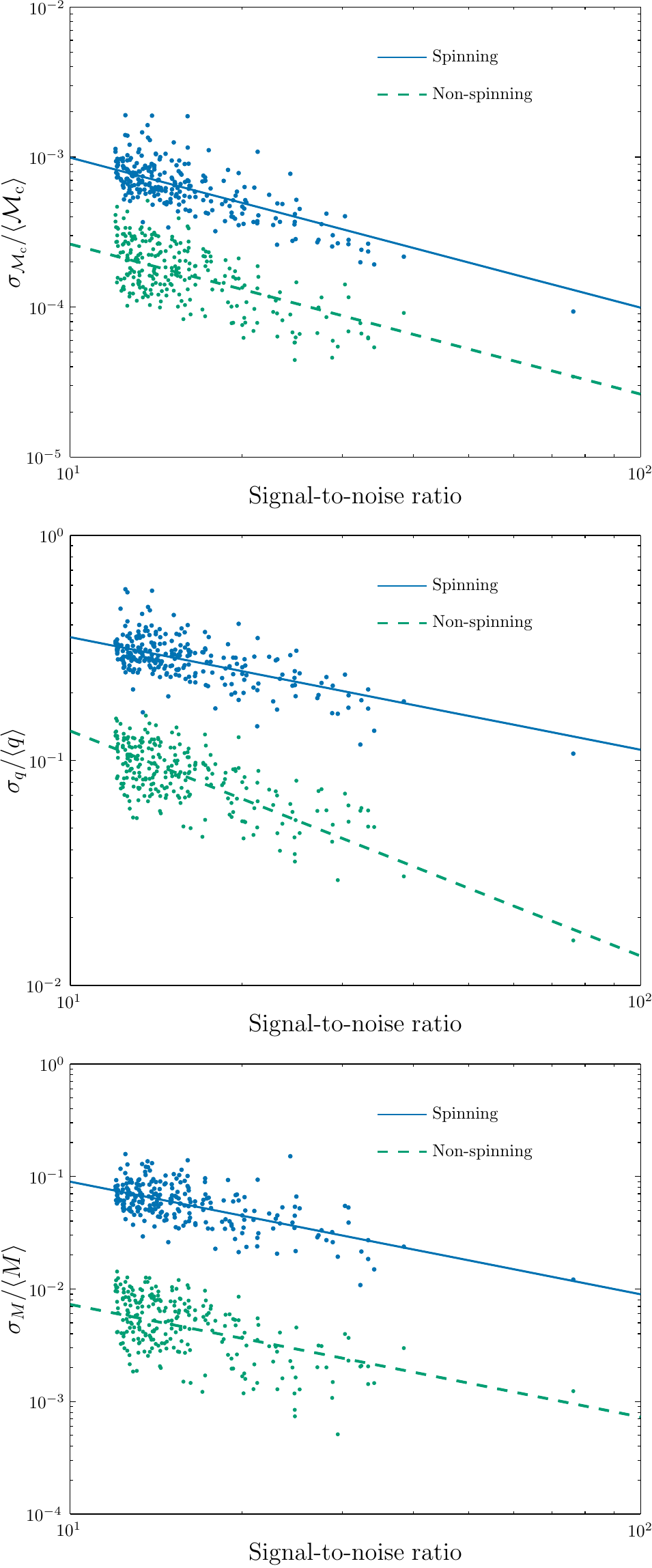}
  \caption{\protect\input{mass_snr-caption.tex}}
\end{figure}

\subsection{Spin Estimates}\label{sec:spin-magnitudes}
We now look at the constraints placed on the spin of the slowly spinning simulated BNS sources.  Even though the simulations occupy a small fraction of the spin-magnitude prior volume, most posterior distributions span the majority of the prior range. For non-precessing systems, where the orbital plane is stationary with respect to the line-of-sight, varying the spin of the compact objects has a similar effect on the phase evolution of the GW as varying the mass ratio. This results in a strong degeneracy between the two parameters.  Modulation of the GWs from precession of the orbital plane can break this degeneracy \citep{Vecchio_2004,Lang_2006,Vitale_2014,Chatziioannou_2014}; however, only systems with large spins that are misaligned with the orbital angular momentum significantly precess. Non-precessing systems, with either low or aligned spins, suffer the most from this degeneracy as the only information regarding the mass and spin is encoded in the phase of the GW.  The simulated sources in this study fall in the latter category of low spins.  Figure  \ref{fig:spinPDFcred} shows the distribution of Gaussian kernel density estimates of the PDFs for the spin of the most and least massive components, $\chi_1$ and $\chi_2$, respectively.  The labeled regions of figure \ref{fig:spinPDFcred} bound the specified percent of PDFs as a function of spin, where the $90\%$ region, for example, is bounded by the $5$th and $95$th percentiles of the PDFs at each spin value. 

The spin of the more massive component has a larger effect on the GW, and is therefore systematically better constrained, as seen in Figure \ref{fig:spinPDFcred}.  For both spins, however, the posterior shows slow spins to be only slightly more probable than high spins for most sources. The mean (median) $50\%$ upper limits on $\chi_1$ and $\chi_2$ are $0.319$ ($0.302$) and $0.424$ ($0.419$) respectively; the $90\%$ upper limits are $0.707$ ($0.699$) and $0.855$ ($0.859$).

\begin{figure}
  \centering
  \includegraphics[width=0.95\columnwidth]{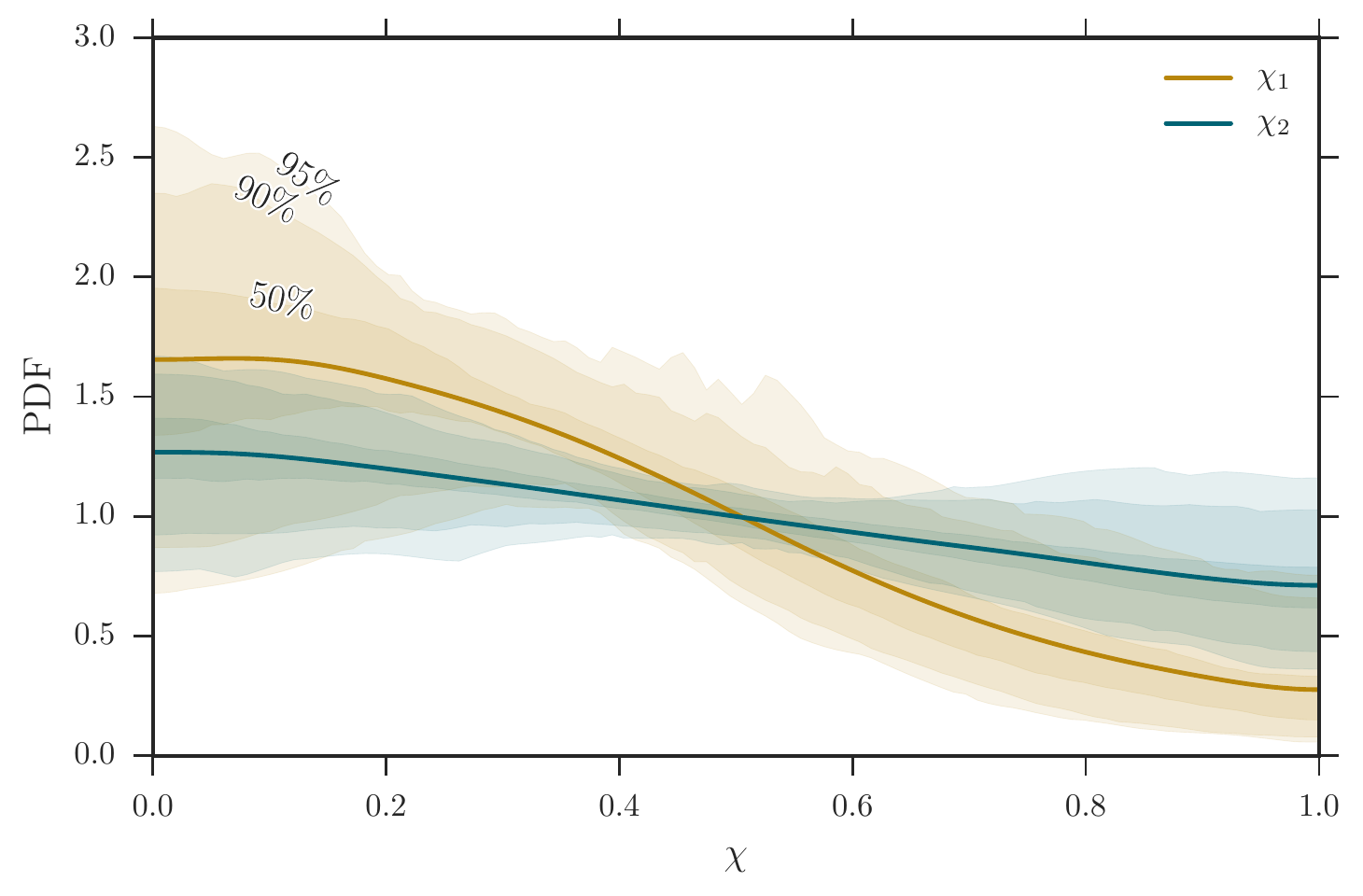}
  \caption{\protect\input{spin_cred_regions-caption.tex}}
\end{figure}

\subsection{Prior Constraints on Spin}
\label{subsec:prior_constraints}

Since spin is largely degenerate with mass ratio, and spin is expected to be small for BNS sources, it is interesting to ask how the mass constraints are affected by making stronger prior assumptions about the spin of NSs.  First, we make the extreme assumption that NSs have negligible spin, as was done in \citet{Singer_2014} and \citet{Berry_2014}.  Figure \ref{fig:mass_std} compares the distribution of (fractional) uncertainties in chirp-mass, mass-ratio and total-mass estimates for the spinning and non-spinning analyses. The average fractional uncertainties from the non-spinning analysis are a factor of $\sim3$--$4$ smaller than the uncertainties from a spinning analysis in the case of $\mathcal{M}_\mathrm{c}$ and $q$, and about an order of magnitude smaller for the total mass.

\begin{figure}
  \centering
  \includegraphics[width=0.85\columnwidth]{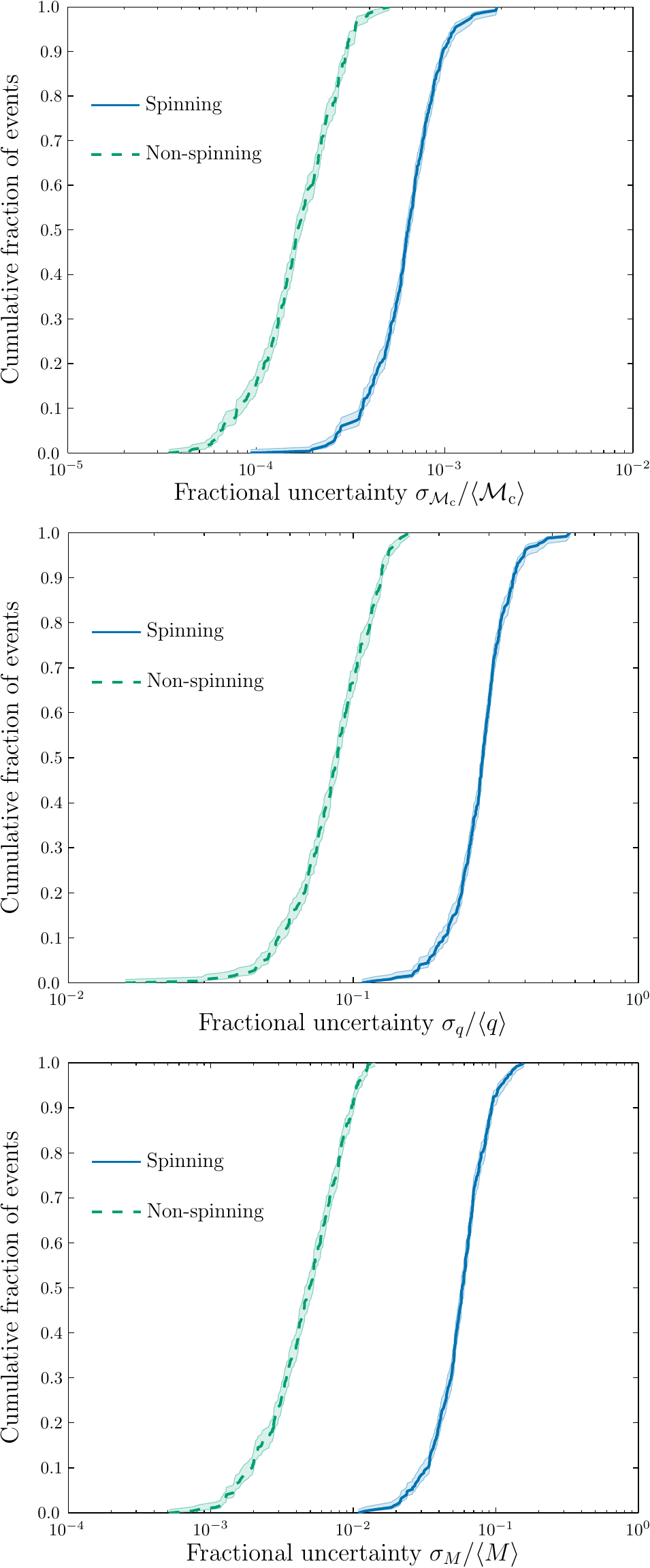}
  \caption{\protect\input{mass_fractional-caption.tex}} 
\end{figure}

Figure \ref{fig:comp_masses} compares cartoon $90\%$ credible regions in component-mass space of $5$ chosen simulated signals \cite[cf.]{Hannam_2013,Chatziioannou_2014}.  As a consequence of the difficulty of estimating the narrow and nonlinearly correlated credible regions in $m_1$--$m_2$ space, we illustrate the credible regions in $m_1$--$m_2$ space as the projection of a rectangular region in $\mathcal{M}_\mathrm{c}$--$q$ space.  To define the rectangular region we use $90\%$ credible intervals of the one-dimensional posterior PDFs of $\mathcal{M}_\mathrm{c}$ and $q$; for $\mathcal{M}_\mathrm{c}$ we use the central $90\%$ credible interval ($5$th to $95$th percentile), and for $q$ the upper $90\%$ credible interval ($10$th to $100$th percentile).  These differing credible intervals were chosen to better summarize the one-dimensional posterior PDFs, which are typically normal for $\mathcal{M}_\mathrm{c}$ and skewed toward high values for $q$ (see figure \ref{fig:mass_pdfs}).

\begin{figure}
  \centering
  \includegraphics[width=1.0\columnwidth]{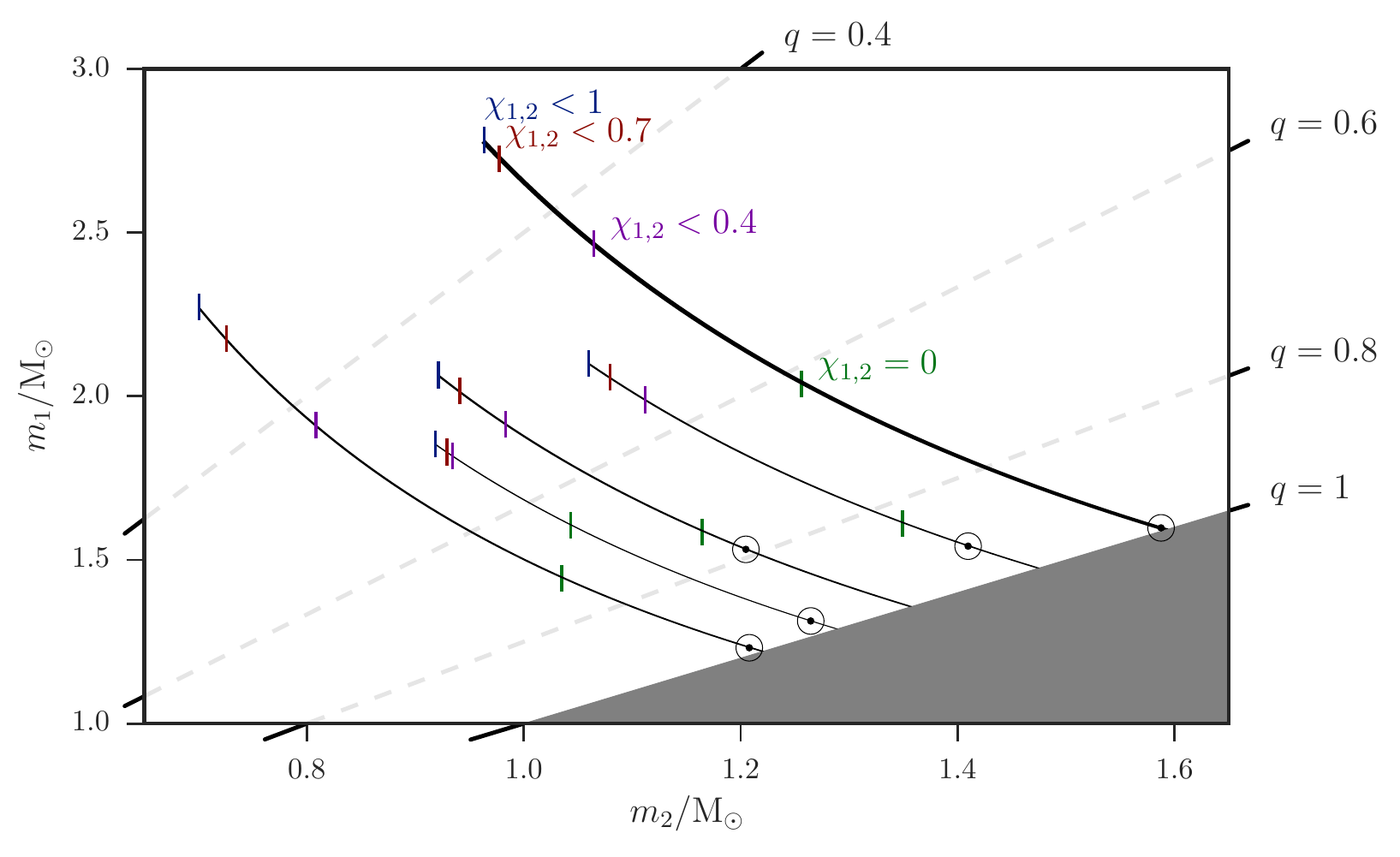}
  \caption{\protect\input{mass_comp-caption.tex}}
\end{figure}

We can investigate the impact of stronger prior assumptions regarding the maximum spin of NSs on mass estimates by discarding posterior samples above a given spin.  Figure \ref{fig:restricted_priors} shows the cumulative distribution of lower $90\%$ bounds on the estimates of $m_2$ among the $250$ simulated sources for spin priors of $\chi_{1,~2} \leq \{1, 0.7, 0.4, 0\}$.  $\chi_{1,~2}<1$ and $\chi_{1,~2}=0$ correspond to the spinning and non-spinning analyses described above.  $\chi<0.7$ is consistent with the NSs remaining intact for most proposed non-exotic EOSs.  $\chi<0.4$ is consistent with the spin of observed, isolated NSs to date.

From these PDFs, it is clear that fairly strong prior assumptions on NS spin are required to significantly impact mass constraints. Assuming NSs to be spinning with $\chi_{1,~2}\leq 0.4$ a priori only constrains masses by an extra few percent compared to allowing them to have $\chi_{1,~2} \leq 1$.

\begin{figure}
  \centering
  \includegraphics[width=0.95\columnwidth]{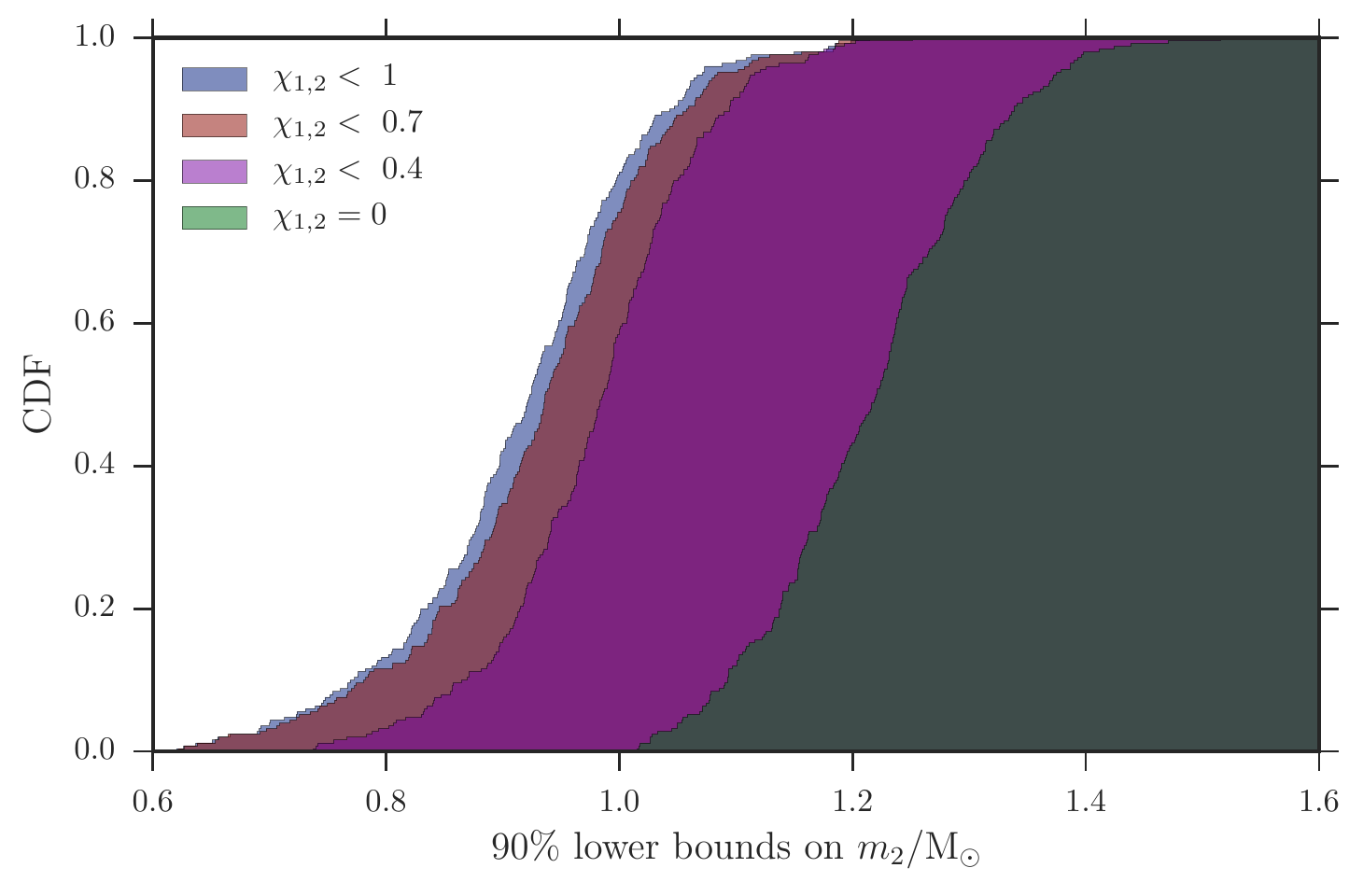}
  \caption{\protect\input{mass_secondary-caption.tex}}
\end{figure}

\section{Source Location}\label{sec:extrinsic}

Having discussed how GW observations can measure the intrinsic properties of their source systems, we now consider the measurement of extrinsic parameters, specifically the sky position (section \ref{sec:sky}) and the distance (section \ref{sec:distance}). These are central to the success of multimessenger astronomy. The sky position is required in order to direct telescopes for electromagnetic (EM) follow-up and to verify that any observed transients do coincide with the source of the gravitational waves. The distance also aids electromagnetic follow-up as it allows cross-reference with galaxy catalogs to find the most probable source locations \citep{Nissanke_2012,Hanna:2013,Fan_2014,Blackburn:2014rqa}. Even without an observed counterpart, the posterior for the (three-dimensional) position allows us to assign a probability that the source resides in given galaxies; combining the redshift of these galaxies (measured electromagnetically) with the gravitational-wave luminosity distance gives a measure of the Hubble constant free of the usual systematics \citep{Schutz_1986,Del_Pozzo_2012}. For our population of slowly spinning NSs, we do not expect the measurement of the extrinsic parameters to be affected by the inclusion of spin in the analysis.

\subsection{Sky Localization}\label{sec:sky}

In order for EM observatories to follow-up a GW detection, they need an accurate sky location. This must be provided promptly, while there is still a visible transient. Parameter estimation while accounting for spin is computationally expensive and slow to complete (see appendix \ref{ap:CPU}). There are alternative methods that can provide sky localization more quickly. The most expedient is \textsc{bayestar}, which uses output from the detection pipeline to rapidly compute sky position \citep{Singer:2015ema}. \textsc{bayestar} can compute sky positions with a latency of a few seconds. Between the low-latency \textsc{bayestar} and the high-latency full parameter estimation, there is the medium-latency option of performing non-spinning parameter estimation with computationally cheap TaylorF2 waveforms. This requires hours of wall time to complete analyses, with the exact time depending upon the degree of parallelization. Despite only using information from the detection triggers, rather than full waveforms, it has been shown that \textsc{bayestar} produces sky areas for BNS signals fully consistent with non-spinning parameter estimation results, provided that there was a trigger from all detectors in the network \citep{Singer_2014,Berry_2014,Singer:2015ema}. Having now performed a full spinning analysis, we can compare the results of high-latency parameter estimation with the more expedient methods of inferring sky position.

In Figure~\ref{fig:sky} we show the cumulative distributions of recovered $50\%$ credible regions, $90\%$ credible regions, and searched areas. All three quantities show good agreement across all parameter-estimation techniques. 
For the for the slowly-spinning BNSs considered here, including spin in the analysis does not change the average ability to localize sources on the sky.

\begin{figure}
  \centering
  \includegraphics[width=0.85\columnwidth]{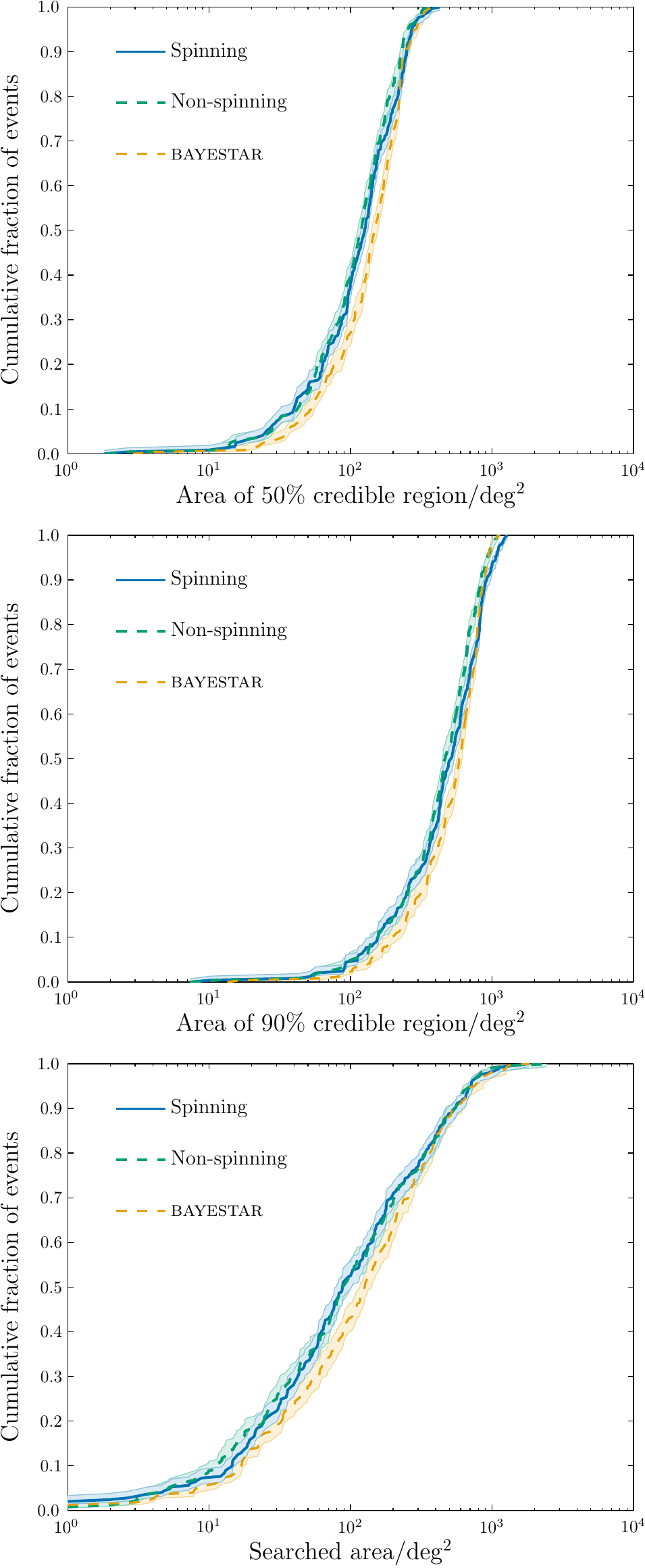}
  \caption{\protect\input{sky_areas-caption.tex}}
\end{figure}

We can consider sky localization in greater detail by comparing areas on an event-by-event basis and not just the cumulative distribution across the population. Doing this, we confirm that sky localization is consistent between approaches for any given event. We use the medium-latency, non-spinning TaylorF2 analysis as a reference point, and compare the ratio of sky areas. To summarize the variation in sky areas computed in different analyses, we use the log ratio
\begin{equation}
\mathcal{R}_A^X = \log_{10}\left(\frac{A^X}{A^\mathrm{NS}}\right),
\end{equation}
where $A^X$ is a credible region or the searched area as determined by method $X$ and $A^\mathrm{NS}$ is the same quantity from the non-spinning analysis. The log ratio $\mathcal{R}_A^X$ is zero when analysis $X$ agrees with the non-spinning results. Considering all $250$ events, the mean and standard deviation of the log ratio is given in Table~\ref{tab:sky-ratio}. For the purposes of EM follow-up, there is no significant difference between analyses.\footnote{The non-spinning analysis was performed with \textsc{LALInference\_nest} while the spinning analysis was performed with \textsc{LALInference\_MCMC} \citep{Veitch_2014}; therefore the consistency between analyses additionally shows the consistency of results from different sampling algorithms.} The computationally expensive fully spinning analysis does not improve sky localization: there is no disadvantage in using the lower-latency results for EM follow-up of slowly spinning BNSs.

\begin{deluxetable}{c cc cc}
\tabletypesize{\scriptsize}
\tablecaption{\label{tab:sky-ratio} Comparison of sky localization areas produced by the low-latency \textsc{bayestar} analysis and the high-latency fully spinning SpinTaylorT4 analysis to those produced by the medium-latency non-spinning TaylorF2 analysis. The mean and standard deviation of the log ratio for the $50\%$ credible region $\mathrm{CR}_{0.5}$, the $90\%$ credible region $\mathrm{CR}_{0.9}$ and the searched area $A_\ast$ are listed for each analysis.}
\tablehead{  
    \colhead{} & \multicolumn{2}{c}{\textsc{bayestar}} & \multicolumn{2}{c}{Spinning} \\
    \colhead{Logarithmic} & \colhead{} & \colhead{Standard} & \colhead{} & \colhead{Standard} \\
    \colhead{ratio} & \colhead{Mean} & \colhead{deviation} & \colhead{Mean} & \colhead{deviation}
}
\startdata
 $\displaystyle \vphantom{\Big|} \mathcal{R}^X_{\mathrm{CR}_{0.5}}$ & $0.095$ & $0.117$ & $0.022$ & $0.067$ \\
 $\displaystyle \vphantom{\Big|} \mathcal{R}^X_{\mathrm{CR}_{0.9}}$ & $0.075$ & $0.094$ & $0.028$ & $0.063$ \\
 $\displaystyle \vphantom{\Big|} \mathcal{R}^X_{A_\ast}$            & $0.106$ & $0.447$ & $0.002$ & $0.397$
\enddata
\end{deluxetable}

\subsection{Luminosity distance}\label{sec:distance}

The distance is degenerate with the inclination \citep{Cutler_1994,Aasi_2013}, and the inclination can be better constrained for precessing systems \citep{van_der_Sluys_2008,Vitale_2014}. Since we are considering a population with low spins, precession is minimal, and there should be little effect from including spin in the analysis.

The absolute size of the distance credible interval $\mathrm{CI}_p^{D}$ approximately scales with the distance, hence we divide the credible interval by the true (injected) distance $D_\star$; this gives an approximate analogue of twice the fractional uncertainty \citep{Berry_2014}. The cumulative distribution of the scaled credible intervals is plotted in Figure \ref{fig:distance}. The mean (median) values of $\mathrm{CI}_{0.5}^{D}/D_\star$ for the spinning and non-spinning analyses are $0.436$ ($0.376$) and $0.426$ ($0.363$) respectively; the values of $\mathrm{CI}_{0.9}^{D}/D_\star$ are $0.981$ ($0.845$) and $0.951$ ($0.819$), and the fractional uncertainties $\sigma_D/\langle D\rangle$ are $0.302$ ($0.262$) and $0.245$ ($0.239$). There is negligible difference between the spinning and non-spinning analyses as expected.

\begin{figure*}
  \centering
  \includegraphics[width=1.7\columnwidth]{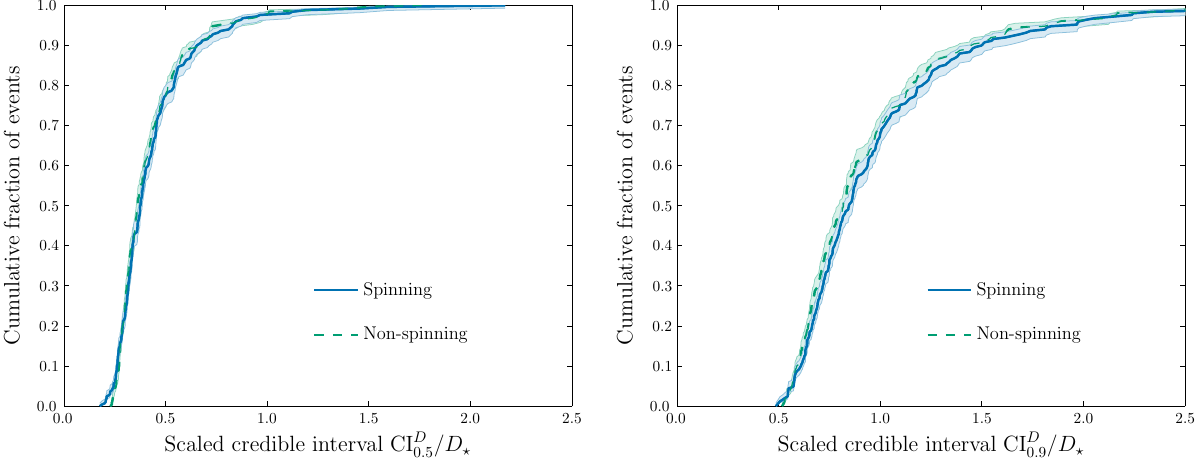}
  \caption{\protect\input{spin_dist-caption.tex}}
\end{figure*}

\section{Conclusions}\label{sec:conclusions}

In this study we investigated the effects of accounting for spin when estimating the parameters of BNS sources with aLIGO. We expect NSs to be only slowly spinning, and hence that their spins only have a small effect of the GW signature of a BNS merger. However, allowing for spins \textit{does} have a significant effect on parameter constraints. Strong degeneracies are present in the model; not only are the spins themselves poorly constrained, but these degeneracies result in weaker constraints on other parameters, particularly masses.  Excluding spin from parameter estimation results in artificially narrow and potentially inaccurate posterior distributions.

Weaker constraints are the result of accounting for broad prior assumptions on NS spins.  We tested various choices for conservative prior assumptions about NS spins and found them to have little effect on mass estimates.  Only strong prior assumptions, such as say $\chi_{1,~2}\lesssim 0.05$ (consistent with the simulated population, and NSs observed in short-period BNS binaries to date) are likely to significantly affect mass constraints.  However, such strict prior assumptions are hard to justify given the small number of observed systems and possible selection effects.

We performed parameter estimation on a astrophysically motivated population of BNS signals, assuming an aLIGO sensitivity comparable to that expected \textbf{throughout} its first observing run. Using a prior on spin magnitudes that is uniform from $0$ to $1$, spanning the range permitted for BHs and extending beyond the expected (but uncertain) upper limit for NSs, the median $90\%$ upper limit on the spin of the more massive component is $0.70$ and the limit for the less massive component is $0.86$. The median fractional uncertainty for the mass ratio $\sigma_q/\langle q \rangle$ is $\sim30\%$, the median fractional uncertainty for the total mass $\sigma_{{M}}/\langle {M} \rangle$ is $\sim6\%$ and the median fractional uncertainty for the chirp mass $\sigma_{\mathcal{M}_\mathrm{c}}/\langle {\mathcal{M}_\mathrm{c}} \rangle$ is $\sim0.06\%$. Despite the mass--spin degeneracy and only weak constraints on the spin magnitudes, we find that we can place precise constraints on the chirp mass for these BNS signals.

The sky-location accuracy, which is central to performing EM follow-up, is not affected by including spin in the analysis of low-spin systems; this may not be the case when spin is higher, i.e.\ in binaries containing a BH. For our population of BNSs, sky localization is unchanged by the inclusion (or exclusion) of spin in parameter estimation. The median $\mathrm{CR}_{0.9}$ ($\mathrm{CR}_{0.5}$) is $\sim 500~\mathrm{deg^2}$ ($\sim 130~\mathrm{deg^2}$). The luminosity distance is similarly unaffected for this population of slowly spinning NSs; the median fractional uncertainty $\sigma_D/\langle D \rangle$ is $\sim 25\%$.  However, an analysis that includes spins requires the use of more computationally expensive waveforms (that include more physics), increasing latency by an order of magnitude.  Therefore, if the population matches our current expectation of being slowly spinning, the low-latency results that could be supplied in time for EM observatories to search for a counter-part are as good as the high-latency results in this respect, and there is no benefit in waiting.

Following the submission of this article, aLIGO made its first detection \citep{Abbott:2016blz}. This was of a binary BH system \citep{TheLIGOScientific:2016wfe} rather than a BNS, but much of our understanding of the abilities of the parameter-estimation analysis, such as the effects of mass--spin degeneracy, translates between sources. The era of GW astronomy has begun, and parameter estimation will play a central role in the science to come.

\acknowledgements

The authors are grateful for useful suggestions from the CBC group of the LIGO Scientific--Virgo Collaboration, in particular we are grateful to Walter Del Pozzo for useful advice, Vivien Raymond for helpful discussions, Simon Stevenson for careful reading and Neil Cornish for beneficial comments.

BF was supported by the Enrico Fermi Institute at the University of Chicago as a McCormick Fellow.  This work was supported in part by the Science and Technology Facilities Council. PBG acknowledges NASA grant NNX12AN10G. SV acknowledges the support of the National Science Foundation and the LIGO Laboratory. JV was supported by STFC grant ST/K005014/1. LIGO was constructed by the California Institute of Technology and Massachusetts Institute of Technology with funding from the National Science Foundation and operates under cooperative agreement PHY-0757058.

This work used computing resources at CIERA funded by NSF PHY-1126812, as well as the computing facilities of the LIGO Data Grid including: the Nemo computing cluster at the Center for Gravitation and Cosmology at the University of Wisconsin--Milwauke under NSF Grants PHY-0923409 and PHY-0600953; the Atlas computing cluster at the Albert Einstein Institute, Hannover; the LIGO computing clusters at Caltech, and the facilities of the Advanced Research Computing @ Cardiff (ARCCA) Cluster at Cardiff University.

Some results were produced using the post-processing tools of the \texttt{plotutils} library at \url{http://github.com/farr/plotutils} and \texttt{skyarea} library at \url{https://github.com/farr/skyarea}.

This paper has been assigned LIGO document reference LIGO-P1500117.

\appendix

\section{Computational cost}\label{ap:CPU}

Performing a fully spinning analysis is computationally expensive. The main computational cost is generating the SpinTaylorT4 waveform, which must be done each time the likelihood is evaluated at a different point in parameter space. Progress is being made in reducing the cost of generating waveforms and evaluating the likelihood \citep[e.g.,][]{Canizares_2013,P_rrer_2014}. Employing reduced order modelling can speed up the non-spinning TaylorF2 analysis by a factor of $\sim 30$ \citep{Canizares_2015}. This is still to be done for a waveform that includes the effects of two unaligned spins; however, progress has also been made in constructing frequency domain approximants using shifted uniform asymptotics, which can speed up generation of a waveform like SpinTaylorT4 by an order of magnitude \citep{Klein:2014bua}.

In figure \ref{fig:wall-time}, we present the approximate wall time taken for analyses comparable to those presented here. The low-latency \textsc{bayestar} and the high-latency fully spinning SpinTaylorT4 results are for the $250$ events considered here. The medium-latency non-spinning TaylorF2 results are from \citet{Berry_2014}; these are not for a different set of signals, but represent a similar population (in more realistic non-Gaussian noise), representing what we hope to achieve in reality.\footnote{We use the more reliably estimated figures for the \textsc{LALInference} runs.} The wall times for \textsc{bayestar} are significantly reduced compared to those in \citet{Berry_2014} because of recent changes to how \textsc{bayestar} integrates over distance \citep{Singer:2015ema}: the mean (median) time is $4.6~\mathrm{s}$ ($4.5~\mathrm{s}$) and the maximum is $6.6~\mathrm{s}$.  We assume that $2000$ (independent) posterior samples are collected for both of the \textsc{LALInference} analyses. The number of samples determines how well we can characterize the posterior: $\sim2000$ is typically needed to calculate $\mathrm{CR}_{0.9}$ to $10\%$ accuracy \citep{DelPozzo_2015}. In practice, we may want to collect additional samples to ensure our results are accurate, but preliminary results could also be released when the medium-latency analysis has collected $1000$ samples, which would after half the time shown here with a maximum wall time of $5.87\times10^4~\mathrm{s} \simeq 16~\mathrm{hr}$. We see that the fully spinning analysis is significantly (here a factor of $\sim20$) more expensive than the non-spinning analysis, taking a mean (median) time of $1.48\times10^6~\mathrm{s} \simeq 17~\mathrm{days}$ ($9.19\times10^5~\mathrm{s} \simeq 11~\mathrm{days}$) and a maximum of $1.48\times10^7~\mathrm{s} \simeq 172~\mathrm{days}$.

The times shown in figure \ref{fig:wall-time} illustrate the hierarchy of times associated with different analyses. However, they should not be used as exact benchmarks for times expected during the first observing run of aLIGO because the version of \textsc{LALInference} used here are not the most up-to-date versions. Following the detection pipeline identifying a candidate BNS signal, we expect \textsc{bayestar} results with latency of a few seconds, non-spinning \textsc{LALInference} results with a latency of a few hours, and fully spinning \textsc{LALInference} results only after weeks of computation. 

While work is underway to improve the latency of and to optimize parameter estimation with \textsc{LALInference}, there is also the possibility of developing new algorithms that provide parameter estimates with lower latency \citep{Haster_2015,Pankow:2015cra}. Improving computational efficiency is important for later observing runs with the advanced-detector network: as sensitivities improve and lower frequencies can be measured, we need to calculate longer waveforms (at even greater expense).

\begin{figure}
  \centering
  \includegraphics[width=0.95\columnwidth]{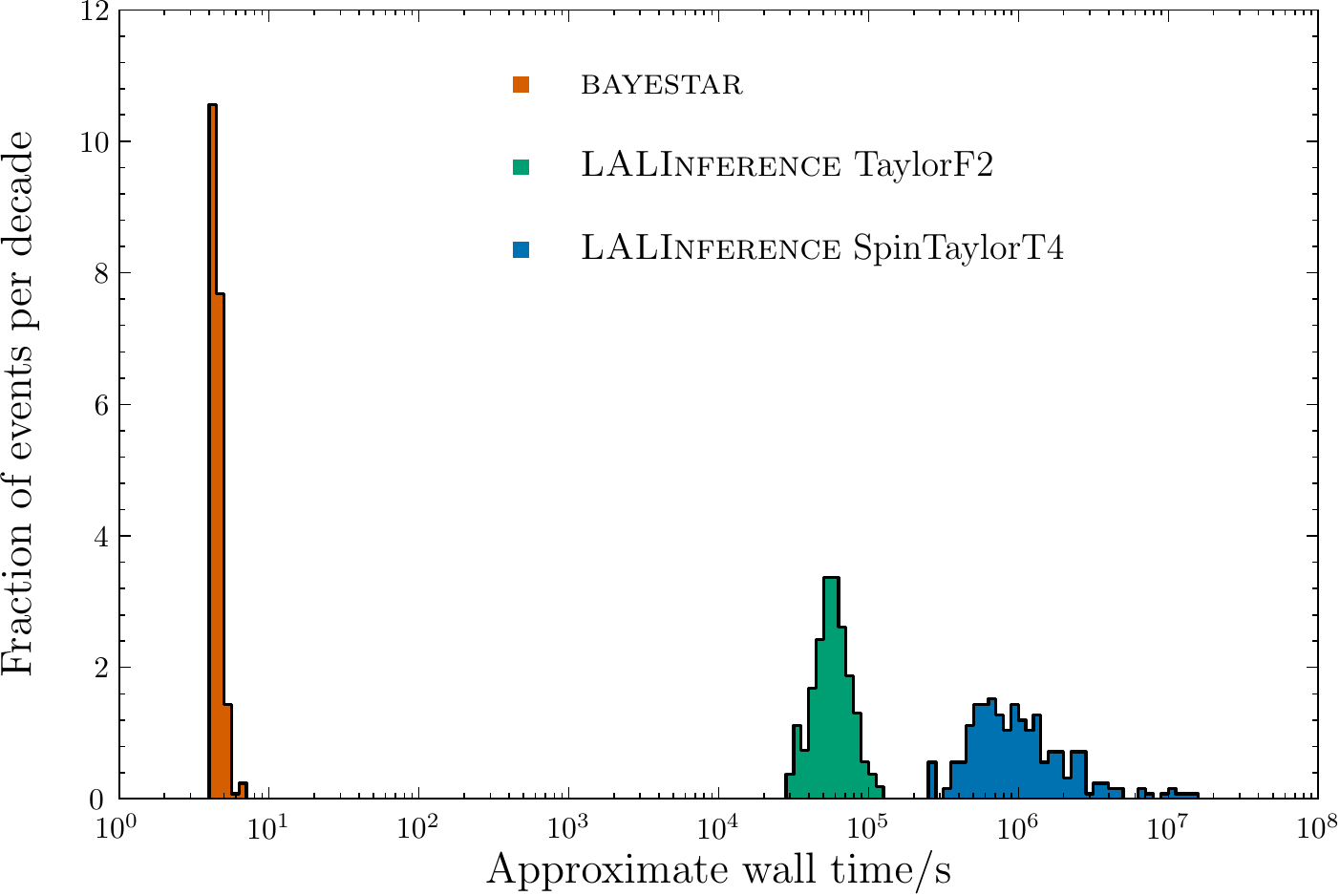}
  \caption{\protect\input{time_hist-caption.tex}}
\end{figure}

\bibliographystyle{apj}
\bibliography{bibliography/biblio} 

\begin{thebibliography}{63}
\providecommand{\natexlab}[1]{#1}
\providecommand{\url}[1]{\texttt{#1}}
\expandafter\ifx\csname urlstyle\endcsname\relax
  \providecommand{\doi}[1]{doi: #1}\else
  \providecommand{\doi}{doi: \begingroup \urlstyle{rm}\Url}\fi

\bibitem[{Aasi} et~al.(2013){Aasi}, {Abadie}, {Abbott}, {Abbott}, {Abbott},
  {Abernathy}, {Accadia}, {Acernese}, {Adams}, {Adams}, and et~al.]{Aasi_2013}
J.~{Aasi}, J.~{Abadie}, B.~P. {Abbott}, R.~{Abbott}, T.~D. {Abbott},
  M.~{Abernathy}, T.~{Accadia}, F.~{Acernese}, C.~{Adams}, T.~{Adams}, and
  et~al.
\newblock {Parameter estimation for compact binary coalescence signals with the
  first generation gravitational-wave detector network}.
\newblock \emph{\prd}, 88:\penalty0 062001, 2013.
\newblock \doi{10.1103/PhysRevD.88.062001}.

\bibitem[{Aasi} et~al.(2015){Aasi}, {Abbott}, {Abbott}, {Abbott}, {Abernathy},
  {Ackley}, {Adams}, {Adams}, {Addesso}, and et~al.]{Aasi_2015}
J.~{Aasi}, B.~P. {Abbott}, R.~{Abbott}, T.~{Abbott}, M.~R. {Abernathy},
  K.~{Ackley}, C.~{Adams}, T.~{Adams}, P.~{Addesso}, and et~al.
\newblock {Advanced LIGO}.
\newblock \emph{Class. Quantum Grav.}, 32:\penalty0 074001, 2015.
\newblock \doi{10.1088/0264-9381/32/7/074001}.

\bibitem[Aasi et~al.(2014)]{WhitePaper2014}
J~Aasi et~al.
\newblock {The LSC--Virgo White Paper on Gravitational Wave Searches and
  Astrophysics}.
\newblock Technical Report LIGO-T1400054-v7, nov 2014.
\newblock URL \url{https://dcc.ligo.org/T1400054/public}.

\bibitem[Abbott et~al.(2016{\natexlab{a}})]{2013arXiv1304.0670L}
B.~P. Abbott et~al.
\newblock {Prospects for Observing and Localizing Gravitational-Wave Transients
  with Advanced LIGO and Advanced Virgo}.
\newblock \emph{Living Rev. Relat.}, 19:\penalty0 1, 2016{\natexlab{a}}.
\newblock \doi{10.1007/lrr-2016-1}.
\newblock URL \url{http://arxiv.org/abs/1304.0670}.

\bibitem[Abbott et~al.(2016{\natexlab{b}})]{Abbott:2016blz}
B.~P. Abbott et~al.
\newblock {Observation of Gravitational Waves from a Binary Black Hole Merger}.
\newblock \emph{\prl}, 116\penalty0 (6):\penalty0 061102, 2016{\natexlab{b}}.
\newblock \doi{10.1103/PhysRevLett.116.061102}.

\bibitem[Abbott et~al.(2016{\natexlab{c}})]{TheLIGOScientific:2016wfe}
B.~P. Abbott et~al.
\newblock {Properties of the binary black hole merger GW150914}.
\newblock 2016{\natexlab{c}}.
\newblock URL \url{http://arxiv.org/abs/1602.03840}.

\bibitem[{Acernese} et~al.(2015){Acernese}, {Agathos}, {Agatsuma}, {Aisa},
  {Allemandou}, {Allocca}, {Amarni}, {Astone}, {Balestri}, {Ballardin}, and
  et~al.]{Acernese_2014}
F.~{Acernese}, M.~{Agathos}, K.~{Agatsuma}, D.~{Aisa}, N.~{Allemandou},
  A.~{Allocca}, J.~{Amarni}, P.~{Astone}, G.~{Balestri}, G.~{Ballardin}, and
  et~al.
\newblock {Advanced Virgo: a second-generation interferometric gravitational
  wave detector}.
\newblock \emph{Class. Quantum Grav.}, 32\penalty0 (2):\penalty0 024001, 2015.
\newblock \doi{10.1088/0264-9381/32/2/024001}.

\bibitem[Barsotti and Fritschel(2012)]{Barsotti:2012}
L.~Barsotti and P.~Fritschel.
\newblock {Early aLIGO Configurations: example scenarios toward design
  sensitivity}.
\newblock Technical Report LIGO-T1200307-v4, aug 2012.
\newblock URL \url{https://dcc.ligo.org/LIGO-T1200307-v4}.

\bibitem[{Berry} et~al.(2015){Berry}, {Mandel}, {Middleton}, {Singer}, {Urban},
  {Vecchio}, {Vitale}, {Cannon}, {Farr}, {Farr}, {Graff}, {Hanna}, {Haster},
  {Mohapatra}, {Pankow}, {Price}, {Sidery}, and {Veitch}]{Berry_2014}
C.~P.~L. {Berry}, I.~{Mandel}, H.~{Middleton}, L.~P. {Singer}, A.~L. {Urban},
  A.~{Vecchio}, S.~{Vitale}, K.~{Cannon}, B.~{Farr}, W.~M. {Farr}, P.~B.
  {Graff}, C.~{Hanna}, C.-J. {Haster}, S.~{Mohapatra}, C.~{Pankow}, L.~R.
  {Price}, T.~{Sidery}, and J.~{Veitch}.
\newblock {Parameter estimation for binary neutron-star coalescences with
  realistic noise during the Advanced LIGO era}.
\newblock \emph{\apj}, 804\penalty0 (2):\penalty0 114, 2015.
\newblock \doi{10.1088/0004-637X/804/2/114}.

\bibitem[{Blackburn} et~al.(2015){Blackburn}, {Briggs}, {Camp}, {Christensen},
  {Connaughton}, {Jenke}, {Remillard}, and {Veitch}]{Blackburn:2014rqa}
L.~{Blackburn}, M.~S. {Briggs}, J.~{Camp}, N.~{Christensen}, V.~{Connaughton},
  P.~{Jenke}, R.~A. {Remillard}, and J.~{Veitch}.
\newblock {High-energy electromagnetic offline follow-up of LIGO-Virgo
  gravitational-wave binary coalescence candidate events}.
\newblock \emph{\apjs}, 217\penalty0 (1):\penalty0 8, 2015.
\newblock \doi{10.1088/0067-0049/217/1/8}.

\bibitem[Brown et~al.(2012)Brown, Harry, Lundgren, and Nitz]{Brown_2012}
Duncan~A. Brown, Ian Harry, Andrew Lundgren, and Alexander~H. Nitz.
\newblock {Detecting binary neutron star systems with spin in advanced
  gravitational-wave detectors}.
\newblock \emph{\prd}, 86:\penalty0 084017, 2012.
\newblock \doi{10.1103/PhysRevD.86.084017}.

\bibitem[Buonanno et~al.(2003)Buonanno, Chen, and Michele]{Buonanno_2003}
Alessandra Buonanno, Yan-bei Chen, and Vallisneri Michele.
\newblock {Detecting gravitational waves from precessing binaries of spinning
  compact objects: Adiabatic limit}.
\newblock \emph{\prd}, 67:\penalty0 104025, 2003.
\newblock \doi{10.1103/PhysRevD.67.104025, 10.1103/PhysRevD.74.029904}.
\newblock [Erratum: Phys. Rev.D 74,029904 (2006)].

\bibitem[Buonanno et~al.(2009)Buonanno, Iyer, Ochsner, Pan, and
  Sathyaprakash]{Buonanno_2009}
Alessandra Buonanno, Bala Iyer, Evan Ochsner, Yi~Pan, and B.~S. Sathyaprakash.
\newblock {Comparison of post-Newtonian templates for compact binary inspiral
  signals in gravitational-wave detectors}.
\newblock \emph{\prd}, 80:\penalty0 084043, 2009.
\newblock \doi{10.1103/PhysRevD.80.084043}.

\bibitem[{Burgay} et~al.(2003){Burgay}, {D'Amico}, {Possenti}, {Manchester},
  {Lyne}, {Joshi}, {McLaughlin}, {Kramer}, {Sarkissian}, {Camilo}, {Kalogera},
  {Kim}, and {Lorimer}]{Burgay_2003}
M.~{Burgay}, N.~{D'Amico}, A.~{Possenti}, R.~N. {Manchester}, A.~G. {Lyne},
  B.~C. {Joshi}, M.~A. {McLaughlin}, M.~{Kramer}, J.~M. {Sarkissian},
  F.~{Camilo}, V.~{Kalogera}, C.~{Kim}, and D.~R. {Lorimer}.
\newblock {An increased estimate of the merger rate of double neutron stars
  from observations of a highly relativistic system}.
\newblock \emph{\nat}, 426:\penalty0 531--533, 2003.
\newblock \doi{10.1038/nature02124}.

\bibitem[{Canizares} et~al.(2015){Canizares}, {Field}, {Gair}, {Raymond},
  {Smith}, and {Tiglio}]{Canizares_2015}
P.~{Canizares}, S.~E. {Field}, J.~{Gair}, V.~{Raymond}, R.~{Smith}, and
  M.~{Tiglio}.
\newblock {Accelerated gravitational-wave parameter estimation with reduced
  order modeling}.
\newblock \emph{\prl}, 114\penalty0 (7):\penalty0 071104, 2015.
\newblock \doi{10.1103/PhysRevLett.114.071104}.
\newblock URL \url{http://dx.doi.org/10.1103/PhysRevLett.114.071104}.

\bibitem[Canizares et~al.(2013)Canizares, Field, R., and
  Tiglio]{Canizares_2013}
Priscilla Canizares, Scott~E. Field, Gair~Jonathan R., and Manuel Tiglio.
\newblock {Gravitational wave parameter estimation with compressed likelihood
  evaluations}.
\newblock \emph{\prd}, 87\penalty0 (12):\penalty0 124005, 2013.
\newblock \doi{10.1103/PhysRevD.87.124005}.
\newblock URL \url{http://dx.doi.org/10.1103/PhysRevD.87.124005}.

\bibitem[{Cannon} et~al.(2012){Cannon}, {Cariou}, {Chapman}, {Crispin-Ortuzar},
  {Fotopoulos}, {Frei}, {Hanna}, {Kara}, {Keppel}, {Liao}, {Privitera},
  {Searle}, {Singer}, and {Weinstein}]{Cannon_2012}
K.~{Cannon}, R.~{Cariou}, A.~{Chapman}, M.~{Crispin-Ortuzar}, N.~{Fotopoulos},
  M.~{Frei}, C.~{Hanna}, E.~{Kara}, D.~{Keppel}, L.~{Liao}, S.~{Privitera},
  A.~{Searle}, L.~{Singer}, and A.~{Weinstein}.
\newblock {Toward Early-Warning Detection of Gravitational Waves from Compact
  Binary Coalescence}.
\newblock \emph{\apj}, 748:\penalty0 136, 2012.
\newblock \doi{10.1088/0004-637X/748/2/136}.

\bibitem[Chatziioannou et~al.(2015)Chatziioannou, Cornish, Antoine, and
  Yunes]{Chatziioannou_2014}
Katerina Chatziioannou, Neil Cornish, Klein Antoine, and Nicol Yunes.
\newblock {Spin-Precession: Breaking the Black Hole--Neutron Star Degeneracy}.
\newblock \emph{\apjl}, 798\penalty0 (1):\penalty0 L17, 2015.
\newblock \doi{10.1088/2041-8205/798/1/L17}.

\bibitem[Christensen et~al.(2004)Christensen, Libson, and
  Meyer]{Christensen_2003}
N.~Christensen, A.~Libson, and R.~Meyer.
\newblock {A Metropolis-Hastings routine for estimating parameters from compact
  binary inspiral events with laser interferometric gravitational radiation
  data}.
\newblock \emph{Class. Quantum Grav.}, 21:\penalty0 317--330, 2004.
\newblock \doi{10.1088/0264-9381/21/1/023}.

\bibitem[Cutler and Flanagan(1994)]{Cutler_1994}
Curt Cutler and Eanna~\'E. Flanagan.
\newblock {Gravitational waves from merging compact binaries: How accurately
  can one extract the binary's parameters from the inspiral wave form?}
\newblock \emph{\prd}, 49:\penalty0 2658--2697, 1994.
\newblock \doi{10.1103/PhysRevD.49.2658}.

\bibitem[Damour et~al.(2012)Damour, Nagar, and Villain]{Damour_2012}
Thibault Damour, Alessandro Nagar, and Loic Villain.
\newblock {Measurability of the tidal polarizability of neutron stars in
  late-inspiral gravitational-wave signals}.
\newblock \emph{\prd}, 85:\penalty0 123007, 2012.
\newblock \doi{10.1103/PhysRevD.85.123007}.

\bibitem[Del~Pozzo(2012)]{Del_Pozzo_2012}
Walter Del~Pozzo.
\newblock {Inference of the cosmological parameters from gravitational waves:
  application to second generation interferometers}.
\newblock \emph{\prd}, 86:\penalty0 043011, 2012.
\newblock \doi{10.1103/PhysRevD.86.043011}.

\bibitem[{Del Pozzo} et~al.(2016)]{DelPozzo_2015}
Walter {Del Pozzo} et~al.
\newblock {Techniques for reconstructing sampled probability distributions: An
  application for gravitational-wave observations of binary neutron stars}.
\newblock Technical Report LIGO-P1500176, 2016.
\newblock URL \url{https://dcc.ligo.org/LIGO-P1500176}.

\bibitem[Fan et~al.(2014)Fan, Messenger, and Heng]{Fan_2014}
XiLong Fan, Christopher Messenger, and Ik~Siong Heng.
\newblock {A Bayesian approach to multi-messenger astronomy: Identification of
  gravitational-wave host galaxies}.
\newblock \emph{\apj}, 795\penalty0 (1):\penalty0 43, 2014.
\newblock \doi{10.1088/0004-637X/795/1/43}.

\bibitem[Farr et~al.(2011)Farr, Sravan, Cantrell, Kreidberg, Bailyn, Ilya, and
  Kalogera]{Farr:2010tu}
Will~M. Farr, Niharika Sravan, Andrew Cantrell, Laura Kreidberg, Charles~D.
  Bailyn, Mandel Ilya, and Vicky Kalogera.
\newblock {The Mass Distribution of Stellar-Mass Black Holes}.
\newblock \emph{\apj}, 741:\penalty0 103, 2011.
\newblock \doi{10.1088/0004-637X/741/2/103}.

\bibitem[Finn and Chernoff(1993)]{FinnChernoff}
Lee~Samuel Finn and David~F. Chernoff.
\newblock {Observing binary inspiral in gravitational radiation: One
  interferometer}.
\newblock \emph{\prd}, 47:\penalty0 2198--2219, Mar 1993.
\newblock \doi{10.1103/PhysRevD.47.2198}.

\bibitem[Gregory(2005)]{Gregory2005}
Philip~C. Gregory.
\newblock \emph{{Bayesian Logical Data Analysis for the Physical Sciences}}.
\newblock Cambridge University Press, Cambridge, 2005.
\newblock ISBN 052184150X.

\bibitem[Hanna et~al.(2014)Hanna, Mandel, and Vousden]{Hanna:2013}
Chad Hanna, Ilya Mandel, and Will Vousden.
\newblock {Utility of galaxy catalogs for following up gravitational waves from
  binary neutron star mergers with wide-field telescopes}.
\newblock \emph{\apj}, 784:\penalty0 8, 2014.
\newblock \doi{10.1088/0004-637X/784/1/8}.

\bibitem[Hannam et~al.(2013)Hannam, Brown, Fairhurst, Fryer, and
  Harry]{Hannam_2013}
Mark Hannam, Duncan~A. Brown, Stephen Fairhurst, Chris~L. Fryer, and Ian~W.
  Harry.
\newblock {{WHEN} {CAN} {GRAVITATIONAL}-{WAVE} {OBSERVATIONS} {DISTINGUISH}
  {BETWEEN} {BLACK} {HOLES} {AND} {NEUTRON} {STARS}?}
\newblock \emph{{ApJ}}, 766\penalty0 (1):\penalty0 L14, mar 2013.
\newblock \doi{10.1088/2041-8205/766/1/l14}.
\newblock URL \url{http://dx.doi.org/10.1088/2041-8205/766/1/L14}.

\bibitem[Haster et~al.(2015)Haster, Mandel, and Farr]{Haster_2015}
Carl-Johan Haster, Ilya Mandel, and Will~M. Farr.
\newblock {Efficient method for measuring the parameters encoded in a
  gravitational-wave signal}.
\newblock \emph{Class. Quantum Grav.}, 32\penalty0 (23):\penalty0 235017, 2015.
\newblock \doi{10.1088/0264-9381/32/23/235017}.
\newblock URL \url{http://arxiv.org/abs/1502.05407}.

\bibitem[Hessels et~al.(2006)Hessels, Ransom, H., Freire, M., and
  Camilo]{Hessels_2006}
Jason W.~T. Hessels, Scott~M. Ransom, Stairs~Ingrid H., Paulo Cesar~Carvalho
  Freire, Kaspi~Victoria M., and Fernando Camilo.
\newblock {A radio pulsar spinning at 716-hz}.
\newblock \emph{Science}, 311:\penalty0 1901--1904, 2006.
\newblock \doi{10.1126/science.1123430}.

\bibitem[Hinderer et~al.(2010)Hinderer, Lackey, Lang, and Read]{Hinderer_2010}
Tanja Hinderer, Benjamin~D. Lackey, Ryan~N. Lang, and Jocelyn~S. Read.
\newblock {Tidal deformability of neutron stars with realistic equations of
  state and their gravitational wave signatures in binary inspiral}.
\newblock \emph{\prd}, 81:\penalty0 123016, 2010.
\newblock \doi{10.1103/PhysRevD.81.123016}.

\bibitem[Klein et~al.(2014)Klein, Cornish, and Yunes]{Klein:2014bua}
Antoine Klein, Neil Cornish, and Nicolás Yunes.
\newblock {Fast Frequency-domain Waveforms for Spin-Precessing Binary
  Inspirals}.
\newblock \emph{\prd}, 90:\penalty0 124029, 2014.
\newblock \doi{10.1103/PhysRevD.90.124029}.

\bibitem[Kreidberg et~al.(2012)Kreidberg, Bailyn, Farr, and
  Kalogera]{Kreidberg:2012ud}
Laura Kreidberg, Charles~D. Bailyn, Will~M. Farr, and Vassiliki Kalogera.
\newblock {Mass Measurements of Black Holes in X-Ray Transients: Is There a
  Mass Gap?}
\newblock \emph{\apj}, 757:\penalty0 36, 2012.
\newblock \doi{10.1088/0004-637X/757/1/36}.

\bibitem[Lang and Hughes(2006)]{Lang_2006}
Ryan~N. Lang and Scott~A. Hughes.
\newblock {Measuring coalescing massive binary black holes with gravitational
  waves: The Impact of spin-induced precession}.
\newblock \emph{\prd}, 74:\penalty0 122001, 2006.
\newblock \doi{10.1103/PhysRevD.74.122001}.
\newblock [Erratum: Phys. Rev.D 77,109901 (2008)].

\bibitem[Littenberg et~al.(2015)Littenberg, Farr, Coughlin, Kalogera, and
  Holz]{Littenberg:2015tpa}
Tyson~B. Littenberg, Ben Farr, Scott Coughlin, Vicky Kalogera, and Daniel~E.
  Holz.
\newblock {Neutron stars versus black holes: probing the mass gap with
  LIGO/Virgo}.
\newblock \emph{\apjl}, 807\penalty0 (2):\penalty0 L24, 2015.
\newblock \doi{10.1088/2041-8205/807/2/L24}.

\bibitem[Lo and Lin(2011)]{Lo_2011}
Ka-Wai Lo and Lap-Ming Lin.
\newblock {The spin parameter of uniformly rotating compact stars}.
\newblock \emph{\apj}, 728:\penalty0 12, 2011.
\newblock \doi{10.1088/0004-637X/728/1/12}.

\bibitem[Lorimer(2008)]{Lorimer_2008}
D.~R. Lorimer.
\newblock {Binary and Millisecond Pulsars}.
\newblock \emph{Living Rev. Relat.}, 11:\penalty0 8, 2008.
\newblock \doi{10.12942/lrr-2008-8}.

\bibitem[Mandel and O'Shaughnessy(2010)]{Mandel_2010}
Ilya Mandel and Richard O'Shaughnessy.
\newblock {Compact Binary Coalescences in the Band of Ground-based
  Gravitational-Wave Detectors}.
\newblock \emph{Class. Quantum Grav.}, 27:\penalty0 114007, 2010.
\newblock \doi{10.1088/0264-9381/27/11/114007}.

\bibitem[Mandel et~al.(2015)Mandel, Haster, Dominik, and
  Belczynski]{Mandel:2015spa}
Ilya Mandel, Carl-Johan Haster, Michal Dominik, and Krzysztof Belczynski.
\newblock {Distinguishing types of compact-object binaries using the
  gravitational-wave signatures of their mergers}.
\newblock \emph{\mnras}, 450\penalty0 (1):\penalty0 L85--L89, 2015.
\newblock \doi{10.1093/mnrasl/slv054}.

\bibitem[Nissanke et~al.(2010)Nissanke, Holz, Hughes, Dalal, and
  Sievers]{Nissanke_2010}
Samaya Nissanke, Daniel~E. Holz, Scott~A. Hughes, Neal Dalal, and Jonathan~L.
  Sievers.
\newblock {Exploring short gamma-ray bursts as gravitational-wave standard
  sirens}.
\newblock \emph{\apj}, 725:\penalty0 496--514, 2010.
\newblock \doi{10.1088/0004-637X/725/1/496}.

\bibitem[Nissanke et~al.(2011)Nissanke, Sievers, Dalal, and
  Holz]{Nissanke_2011}
Samaya Nissanke, Jonathan Sievers, Neal Dalal, and Daniel Holz.
\newblock {Localizing compact binary inspirals on the sky using ground-based
  gravitational wave interferometers}.
\newblock \emph{\apj}, 739:\penalty0 99, 2011.
\newblock \doi{10.1088/0004-637X/739/2/99}.

\bibitem[Nissanke et~al.(2013)Nissanke, Kasliwal, and Alexandra]{Nissanke_2012}
Samaya Nissanke, Mansi Kasliwal, and Georgieva Alexandra.
\newblock {Identifying Elusive Electromagnetic Counterparts to Gravitational
  Wave Mergers: an end-to-end simulation}.
\newblock \emph{\apj}, 767:\penalty0 124, 2013.
\newblock \doi{10.1088/0004-637X/767/2/124}.

\bibitem[\"Ozel et~al.(2010)\"Ozel, Psaltis, Narayan, and
  McClintock]{Ozel:2010su}
Feryal \"Ozel, Dimitrios Psaltis, Ramesh Narayan, and Jeffrey~E. McClintock.
\newblock {The Black Hole Mass Distribution in the Galaxy}.
\newblock \emph{\apj}, 725:\penalty0 1918--1927, 2010.
\newblock \doi{10.1088/0004-637X/725/2/1918}.

\bibitem[\"Ozel et~al.(2012)\"Ozel, Psaltis, Narayan, and
  Villarreal]{_zel_2012}
Feryal \"Ozel, Dimitrios Psaltis, Ramesh Narayan, and Antonio~Santos
  Villarreal.
\newblock {On the Mass Distribution and Birth Masses of Neutron Stars}.
\newblock \emph{\apj}, 757:\penalty0 55, 2012.
\newblock \doi{10.1088/0004-637X/757/1/55}.

\bibitem[Pankow et~al.(2015)Pankow, Brady, Ochsner, and
  O'Shaughnessy]{Pankow:2015cra}
C.~Pankow, P.~Brady, E.~Ochsner, and R.~O'Shaughnessy.
\newblock {Novel scheme for rapid parallel parameter estimation of
  gravitational waves from compact binary coalescences}.
\newblock \emph{\prd}, 92\penalty0 (2):\penalty0 023002, 2015.
\newblock \doi{10.1103/PhysRevD.92.023002}.

\bibitem[P\"urrer(2014)]{P_rrer_2014}
Michael P\"urrer.
\newblock {Frequency domain reduced order models for gravitational waves from
  aligned-spin compact binaries}.
\newblock \emph{Class. Quantum Grav.}, 31\penalty0 (19):\penalty0 195010, 2014.
\newblock \doi{10.1088/0264-9381/31/19/195010}.

\bibitem[{Rodriguez} et~al.(2014){Rodriguez}, {Farr}, {Raymond}, {Farr},
  {Littenberg}, {Fazi}, and {Kalogera}]{Rodriguez_2014}
C.~L. {Rodriguez}, B.~{Farr}, V.~{Raymond}, W.~M. {Farr}, T.~B. {Littenberg},
  D.~{Fazi}, and V.~{Kalogera}.
\newblock {Basic Parameter Estimation of Binary Neutron Star Systems by the
  Advanced LIGO/Virgo Network}.
\newblock \emph{\apj}, 784:\penalty0 119, 2014.
\newblock \doi{10.1088/0004-637X/784/2/119}.

\bibitem[R\"over et~al.(2006)R\"over, Meyer, and Nelson]{R_ver_2006}
Christian R\"over, Renate Meyer, and Christensen Nelson.
\newblock {Bayesian inference on compact binary inspiral gravitational
  radiation signals in interferometric data}.
\newblock \emph{Class. Quantum Grav.}, 23:\penalty0 4895--4906, 2006.
\newblock \doi{10.1088/0264-9381/23/15/009}.

\bibitem[Schutz(1986)]{Schutz_1986}
Bernard~F. Schutz.
\newblock {Determining the Hubble Constant from Gravitational Wave
  Observations}.
\newblock \emph{\nat}, 323:\penalty0 310--311, 1986.
\newblock \doi{10.1038/323310a0}.

\bibitem[{Sidery} et~al.(2014){Sidery}, {Aylott}, {Christensen}, {Farr},
  {Farr}, {Feroz}, {Gair}, {Grover}, {Graff}, {Hanna}, {Kalogera}, {Mandel},
  {O'Shaughnessy}, {Pitkin}, {Price}, {Raymond}, {R{\"o}ver}, {Singer}, {van
  der Sluys}, {Smith}, {Vecchio}, {Veitch}, and {Vitale}]{Sidery_2014}
T.~{Sidery}, B.~{Aylott}, N.~{Christensen}, B.~{Farr}, W.~{Farr}, F.~{Feroz},
  J.~{Gair}, K.~{Grover}, P.~{Graff}, C.~{Hanna}, V.~{Kalogera}, I.~{Mandel},
  R.~{O'Shaughnessy}, M.~{Pitkin}, L.~{Price}, V.~{Raymond}, C.~{R{\"o}ver},
  L.~{Singer}, M.~{van der Sluys}, R.~J.~E. {Smith}, A.~{Vecchio}, J.~{Veitch},
  and S.~{Vitale}.
\newblock {Reconstructing the sky location of gravitational-wave detected
  compact binary systems: methodology for testing and comparison}.
\newblock \emph{\prd}, 89\penalty0 (8):\penalty0 084060, 2014.
\newblock \doi{10.1103/PhysRevD.89.084060}.

\bibitem[{Singer} et~al.(2014){Singer}, {Price}, {Farr}, {Urban}, {Pankow},
  {Vitale}, {Veitch}, {Farr}, {Hanna}, {Cannon}, {Downes}, {Graff}, {Haster},
  {Mandel}, {Sidery}, and {Vecchio}]{Singer_2014}
L.~P. {Singer}, L.~R. {Price}, B.~{Farr}, A.~L. {Urban}, C.~{Pankow},
  S.~{Vitale}, J.~{Veitch}, W.~M. {Farr}, C.~{Hanna}, K.~{Cannon}, T.~{Downes},
  P.~{Graff}, C.-J. {Haster}, I.~{Mandel}, T.~{Sidery}, and A.~{Vecchio}.
\newblock {The first two years of electromagnetic follow-up with Advanced LIGO
  and Virgo}.
\newblock \emph{\apj}, 795\penalty0 (2):\penalty0 105, 2014.
\newblock \doi{10.1088/0004-637X/795/2/105}.

\bibitem[Singer and Price(2016)]{Singer:2015ema}
Leo~P. Singer and Larry~R. Price.
\newblock {Rapid Bayesian position reconstruction for gravitational-wave
  transients}.
\newblock \emph{\prd}, 93\penalty0 (2):\penalty0 024013, 2016.
\newblock \doi{10.1103/PhysRevD.93.024013}.

\bibitem[Skilling(2006)]{Skilling2006}
J.~Skilling.
\newblock {Nested sampling for general {B}ayesian computation}.
\newblock \emph{Bayesian Analysis}, 1\penalty0 (4):\penalty0 833--860, 2006.
\newblock \doi{10.1214/06-BA127}.

\bibitem[Vallisneri(2008)]{Vallisneri_2008}
Michele Vallisneri.
\newblock {Use and abuse of the Fisher information matrix in the assessment of
  gravitational-wave parameter-estimation prospects}.
\newblock \emph{\prd}, 77:\penalty0 042001, 2008.
\newblock \doi{10.1103/PhysRevD.77.042001}.

\bibitem[{van der Sluys} et~al.(2008){van der Sluys}, {R{\"o}ver}, {Stroeer},
  {Raymond}, {Mandel}, {Christensen}, {Kalogera}, {Meyer}, and
  {Vecchio}]{van_der_Sluys_2008}
M.~V. {van der Sluys}, C.~{R{\"o}ver}, A.~{Stroeer}, V.~{Raymond}, I.~{Mandel},
  N.~{Christensen}, V.~{Kalogera}, R.~{Meyer}, and A.~{Vecchio}.
\newblock {Gravitational-Wave Astronomy with Inspiral Signals of Spinning
  Compact-Object Binaries}.
\newblock \emph{\apjl}, 688:\penalty0 L61, 2008.
\newblock \doi{10.1086/595279}.

\bibitem[Vecchio(2004)]{Vecchio_2004}
Alberto Vecchio.
\newblock {LISA observations of rapidly spinning massive black hole binary
  systems}.
\newblock \emph{\prd}, 70:\penalty0 042001, 2004.
\newblock \doi{10.1103/PhysRevD.70.042001}.

\bibitem[Veitch and Vecchio(2010)]{Veitch_2010}
J.~Veitch and A.~Vecchio.
\newblock {Bayesian coherent analysis of in-spiral gravitational wave signals
  with a detector network}.
\newblock \emph{\prd}, 81:\penalty0 062003, 2010.
\newblock \doi{10.1103/PhysRevD.81.062003}.

\bibitem[{Veitch} et~al.(2012){Veitch}, {Mandel}, {Aylott}, {Farr}, {Raymond},
  {Rodriguez}, {van der Sluys}, {Kalogera}, and {Vecchio}]{Veitch_2012}
J.~{Veitch}, I.~{Mandel}, B.~{Aylott}, B.~{Farr}, V.~{Raymond}, C.~{Rodriguez},
  M.~{van der Sluys}, V.~{Kalogera}, and A.~{Vecchio}.
\newblock {Estimating parameters of coalescing compact binaries with proposed
  advanced detector networks}.
\newblock \emph{\prd}, 85:\penalty0 104045, 2012.
\newblock \doi{10.1103/PhysRevD.85.104045}.

\bibitem[{Veitch} et~al.(2015){Veitch}, {Raymond}, {Farr}, {Farr}, {Graff},
  {Vitale}, {Aylott}, {Blackburn}, {Christensen}, {Coughlin}, {Del Pozzo},
  {Feroz}, {Gair}, {Haster}, {Kalogera}, {Littenberg}, {Mandel},
  {O'Shaughnessy}, {Pitkin}, {Rodriguez}, {R{\"o}ver}, {Sidery}, {Smith}, {Van
  Der Sluys}, {Vecchio}, {Vousden}, and {Wade}]{Veitch_2014}
J.~{Veitch}, V.~{Raymond}, B.~{Farr}, W.~{Farr}, P.~{Graff}, S.~{Vitale},
  B.~{Aylott}, K.~{Blackburn}, N.~{Christensen}, M.~{Coughlin}, W.~{Del Pozzo},
  F.~{Feroz}, J.~{Gair}, C.-J. {Haster}, V.~{Kalogera}, T.~{Littenberg},
  I.~{Mandel}, R.~{O'Shaughnessy}, M.~{Pitkin}, C.~{Rodriguez}, C.~{R{\"o}ver},
  T.~{Sidery}, R.~{Smith}, M.~{Van Der Sluys}, A.~{Vecchio}, W.~{Vousden}, and
  L.~{Wade}.
\newblock {Parameter estimation for compact binaries with ground-based
  gravitational-wave observations using the LALInference software library}.
\newblock \emph{\prd}, 91\penalty0 (4):\penalty0 042003, 2015.
\newblock \doi{10.1103/PhysRevD.91.042003}.

\bibitem[Vitale et~al.(2014)Vitale, Lynch, Veitch, Raymond, and
  Sturani]{Vitale_2014}
Salvatore Vitale, Ryan Lynch, John Veitch, Vivien Raymond, and Riccardo
  Sturani.
\newblock {Measuring the spin of black holes in binary systems using
  gravitational waves}.
\newblock \emph{\prl}, 112\penalty0 (25):\penalty0 251101, 2014.
\newblock \doi{10.1103/PhysRevLett.112.251101}.

\bibitem[{Wade} et~al.(2014){Wade}, {Creighton}, {Ochsner}, {Lackey}, {Farr},
  {Littenberg}, and {Raymond}]{Wade_2014}
L.~{Wade}, J.~D.~E. {Creighton}, E.~{Ochsner}, B.~D. {Lackey}, B.~F. {Farr},
  T.~B. {Littenberg}, and V.~{Raymond}.
\newblock {Systematic and statistical errors in a bayesian approach to the
  estimation of the neutron-star equation of state using advanced gravitational
  wave detectors}.
\newblock \emph{\prd}, 89\penalty0 (10):\penalty0 103012, 2014.
\newblock \doi{10.1103/PhysRevD.89.103012}.

\bibitem[Yagi and Yunes(2014)]{Yagi_2014}
Kent Yagi and Nicolas Yunes.
\newblock {Love can be Tough to Measure}.
\newblock \emph{\prd}, 89\penalty0 (2):\penalty0 021303, 2014.
\newblock \doi{10.1103/PhysRevD.89.021303}.

\end{thebibliography}

\end{document}